\documentclass[fleqn,10pt]{wlscirep}
\usepackage[utf8]{inputenc}
\usepackage[T1]{fontenc}
\usepackage{bm}
\usepackage{graphicx}%
\usepackage{amsthm}%
\usepackage{mathptmx}   
\usepackage[title]{appendix}%
\usepackage{xcolor}%
\usepackage{textcomp,manyfoot,booktabs}%
\usepackage{algorithm,algorithmicx,algpseudocode}%
\usepackage{listings,multicol, multirow,upgreek}
\newcommand{\eg}{\textit{e.g.},~}
\newcommand{\ie}{\textit{i.e.},~}

\usepackage{subcaption}

\title{Signals on Graphs Over Time: Methodologies, Applications, and Challenges}
\author[1$\dagger$]{Yi Yan}
\author[1$\dagger$]{Jiacheng Hou}
\author[1]{Zhenjie Song}
\author[1]{Dayu Qin}
\author[1*]{Ercan Engin Kuruoglu}
\affil[1]{Institute of Data and Information, Shenzhen International Graduate School, Tsinghua University, Shenzhen, China}
\affil[$\dagger$]{Equal contribution}
\affil[*]{E-mail: kuruoglu@sz.tsinghua.edu.cn}

\begin{abstract}
Graphs, as effective representations of multidimensional, non-Euclidean, and irregularly structured data, have attracted significant research attention in recent years and are widely applied in diverse real-world domains such as social networks, traffic systems, and energy infrastructures. However, conventional graph-based techniques postulate that the graph is static in both signal and topology. Given the inherently dynamic nature of many real-world systems, research on Time-Varying Graphs (TVGs) has emerged as a critical area, offering the capability to capture essential dynamics of signal and topology changes over time. In the field of Graph Signal Processing (GSP), foundational concepts such as shifting, convolution, filtering, and transformation are adapted to address the challenges presented by time-varying signals on graphs. These techniques are further extended to enhance the implementation and interpretation of Graph Neural Networks (GNNs). Despite rapid progress, the relationships between GSP formulations and GNN architectures for TVGs remain fragmented. This review surveys recent advances in algorithms, applications, and benchmark datasets for TVGs, providing a unified perspective on how GSP principles inform the design and interpretation of graph neural networks for dynamic networks.
\end{abstract}

\begin{document}
\flushbottom
\maketitle
\thispagestyle{empty}

\section{Introduction}
In the present intelligent era, large-scale data collected from complex systems, such as urban transportation networks, meteorological observations, and brain connectivity, are increasingly represented as graphs and networks \cite{transportation, Spelta_2020_NLMS, Huang_2018_brain_GSP}. This representation enables the analysis of structured and interdependent data that traditional Euclidean models cannot adequately describe. 
Early studies in graph-based data analysis and Graph Signal Processing (GSP) typically assume that the underlying graph structure remains static, allowing classical signal processing concepts to be extended to graph domains, including graph shift, graph convolution, Graph Fourier Transform (GFT), Graph Wavelet Transform (GWT), and various graph filters such as low-pass, high-pass, and band-pass filters~\cite{Shuman_2013_the_emerging, Ortega_graph_2018, HAMMOND2011_wavelet, tremblay_2018_filter_design}. 
However, many real-world systems exhibit inherently dynamic behavior in which both the graph signals and network topology evolve. 
For example, brain connectivity patterns vary across cognitive states \cite{zhao2024sequential}, while transportation and environmental monitoring networks continuously change due to temporal fluctuations in demand and physical conditions \cite{yu2018_STGCN}. 
These dynamics challenge the assumptions of static graph models and motivate the development of signal processing frameworks capable of handling time-varying graphs (TVGs).

Even though treating TVGs as multivariate time series provides a straightforward approach for data processing, this simplification ignores the interactions among observations that are encoded in the underlying graph topology. 
Such interactions are often essential for understanding the dynamics of many real-world systems. 
For example, in missing-not-at-random scenarios where observations associated with certain variables or entities are consistently unavailable, the missing entries may correspond to features that are systematically unobserved or unreliable \cite{Decorte_Missing_2024}. 
In many real-world systems, these variables are not independent but exhibit strong dependencies. Exploiting these relationships allows missing values to be inferred from related observations, whereas approaches that ignore such dependencies struggle to produce reliable estimates.
Moreover, conventional time-series models typically rely on independent or identically distributed assumptions, which fail to capture the relational dependencies naturally represented by graph connectivity. 
These limitations highlight the need for dedicated algorithms and models capable of processing signals on TVGs.

As an extension of GSP, time-varying Graph Signal Processing (TVGSP) has emerged to address dynamic graph data by modeling evolving signals through adaptive updating and spatiotemporal relationship estimation. 
Meanwhile, graph neural networks (GNNs), with their strong capability for nonlinear and complex representation learning, have become powerful tools for processing graph-structured data~\cite{kipf2017semi_supervised, velickovic2018graph_GAT}. 
Building on this framework, time-varying Graph Neural Networks (TVGNNs) have been developed to address dynamic graph scenarios through spatiotemporal learning and continual learning mechanisms~\cite{dong_2016_learning_laplacian, wang2022lifelong}. 
In both TVGSP and TVGNN approaches, a central challenge is the effective modeling of joint variations across spatial and temporal domains. 
TVGSP and TVGNN both focus on modeling and learning from graph-structured data that evolves over time, providing complementary tools for understanding and processing spatiotemporal graph signals.
In particular, TVGSP provides theoretical foundations for understanding the evolution of signals on dynamic graphs, where insights into the behavior of time-varying data can inform the design and interpretation of TVGNN architectures.

\begin{figure*}
    \centering
    \includegraphics[width=\linewidth]{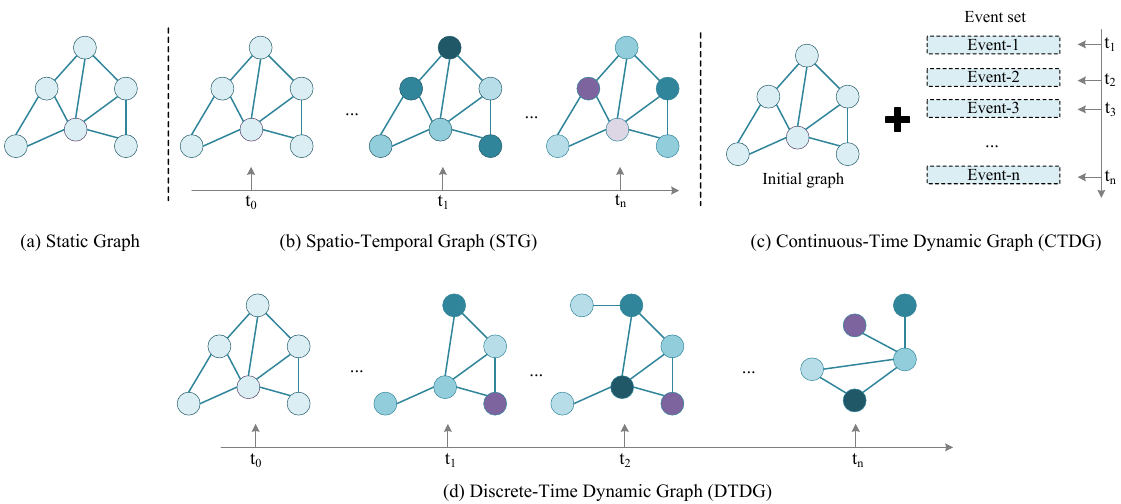}
    \caption{Illustration of different types of graphs}
    \label{fig_graph_types}
\end{figure*}

In this review, we aim to systematically connect TVGSP and TVGNNs by examining their methodologies and analyzing their respective strengths and limitations. 
While studies have reviewed GSP~\cite{Ortega_graph_2018, leus2023graph, Dong_Graph_ML_2020, Isufi_2024_graph} and GNNs~\cite{corso2024graph, wu2020comprehensive,zhou2020graph, Ruiz_2021_graph, Chien2024}, existing surveys primarily focus on static graphs with time-invariant data.
Several recent surveys have provided comprehensive overviews of TVGNNs from deep-learning perspectives, including spatial models and tasks for dynamic graphs~\cite{Gravina_2024_deep_learning, longa2023graph, zheng2025survey}, spatio-temporal GNN architectures~\cite{Cini_2025_graph, sahili2023spatio, jin2023spatio}, and dynamic graph representation learning frameworks~\cite{Yang_2024_dynamic}.
These works offer valuable coverage of TVGNN taxonomies, datasets, and application trends. However, most previous studies examine only one paradigm in isolation, either GSP or GNN, thereby overlooking the fundamental connections between TVGSP and TVGNNs.
In contrast, this review focuses on bridging TVGSP and dynamic GNN models, highlighting how GSP formulations, spectral interpretations, and filtering principles provide theoretical insights that inform, complement, and sometimes extend neural spatiotemporal architectures. 
As closely related approaches for modeling signals on TVGs, a unified review of TVGSP and TVGNNs can offer deeper insights into dynamic graph learning and help guide future research directions.
The main contributions are summarized as follows:



\begin{itemize}
    \item \textbf{Comprehensive framework}: We provide a structured review of the fundamental concepts and defining characteristics of TVGSP and TVGNNs, establishing a cohesive framework that organizes key techniques and representative models.
    \item \textbf{Unified perspective}: We present a unifying perspective that connects mainstream TVGNN approaches with TVGSP principles, clarifying their relationships and discussing advantages, limitations, trade-offs, and practical considerations in real-world TVG applications.
    \item \textbf{Future research directions}: We discuss open challenges and emerging opportunities in processing TVGs, including new perspectives on discrete-time dynamic graphs (DTDG) and continuous-time dynamic graphs (CTDG) within the TVGSP framework, and highlight promising research directions.
\end{itemize}
In Section~\ref{sec_background}, we begin with important definitions and background knowledge about TVGSP and TVGNNs. In Section~\ref{sec_graph_structure}, we focus on the acquisition and representation of TVGs. Sections~\ref{sec_GSP} and~\ref{section_GNN} provide a comprehensive review and discussion of the existing approaches for processing TVGs, including both signal processing and neural network methods. Section~\ref{sec_application} summarizes the commonly used datasets, the corresponding application, and the evaluation metrics. Finally, Section~\ref{sec_conclusion} concludes and indicates the future direction of the TVGSP.

\section{Preliminaries}
\label{sec_background}

\begin{figure*}[h]
    \centering
    \includegraphics[trim={25 585 170 0},clip,width=\linewidth]{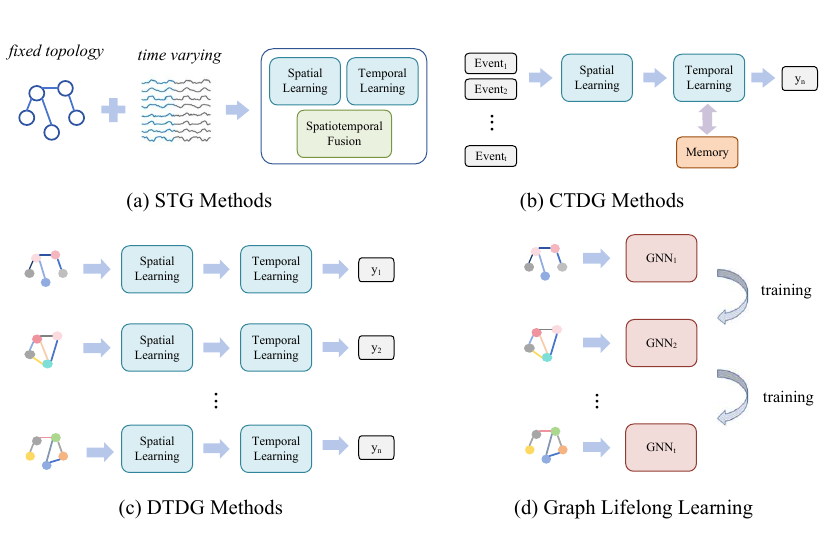}
    \caption{Illustration of different genres of TVG methods.}
    \label{fig:tvgnn}
\end{figure*}

Graphs effectively encapsulate complex interactions within multivariate systems, where pairwise relationships between variables are encoded as edges. The nodes, edges, and the data associated with them can be dynamically updated to reflect the intrinsic behavior and evolving nature of the underlying relationships.
Graph representations emerge and capture the structure and interdependencies of the data.

\subsection{Definition of Time-Varying Graphs and Signals}
In conventional GSP that operates on static data, the main research focus is placed on a static graph $\mathcal{G} = (\mathcal{V}, \mathcal{E})$, which is defined by a set of nodes $\mathcal{V} = {v_1, \dots, v_N}$ and a set of edges $\mathcal{E}$ that represent the connections between nodes. 
In this review, the graph $\mathcal{G}$ is assumed undirected unless otherwise specified, and $\mathcal{G}$ can be either weighted or unweighted. 
A graph signal $\boldsymbol{x}$ is defined as a multi-dimensional vector, recording the value of each node in $\mathcal{G}$. 
We further denote the number of nodes with $|\mathcal{V}| = N$.
The adjacency matrix $\mathbf{A}$ of $\mathcal{G}$ comprises the topological information by storing the state of connectivity between nodes. 
To be specific, the $ij^\text{th}$ entry of $\mathbf{A}$ is the edge weight between nodes $v_i$ and $v_j$. If $\mathcal{G}$ is unweighted, $A_{ij} = 1$ indicates there is an edge between nodes $v_i$ and $v_j$. 
Conversely, $A_{ij} = 0$ means there is no connection between $v_i$ and $v_j$. For an undirected graph, the adjacency matrix $\mathbf{A}$ is always symmetric. 
If $\mathcal{G}$ is undirected and unweighted, the number of edges connected to a node $v_i$ is defined as the node degree $d_i$, which can be represented by a diagonal matrix called the degree matrix $\mathbf{D} = \text{diag}(d_1, \dots, d_N)$. 
In the weighted case, the degree of a node is the sum of all edge weights connected to it, rather than simply counting the number of edges.

Since real-world data often evolves over time, meaning that its graph representations, including node values, edge weights, and even graph topology, could be dynamic, applying static techniques independently at each time instance is inadequate, as it neglects the temporal dependencies inherent in dynamic systems. 
Consequently, a variety of TVG algorithms have been developed to address different forms of graph dynamics.
In this review, following previous works~\cite{jin2023spatio, feng2024comprehensive, zheng2025survey}, we categorized TVGs into three classes from the evolutionary perspective: the Spatiotemporal Graphs (STGs), the Discrete-Time Dynamic Graphs (DTDGs), and the Continuous-Time Dynamic Graphs (CTDGs) as illustrated in Fig.~\ref{fig_graph_types}.
\begin{figure*}[htb]
    \centering
    \begin{subfigure}[t]{0.48\linewidth}
        \centering
        \includegraphics[trim={75 75 15 70},clip,width=\linewidth]{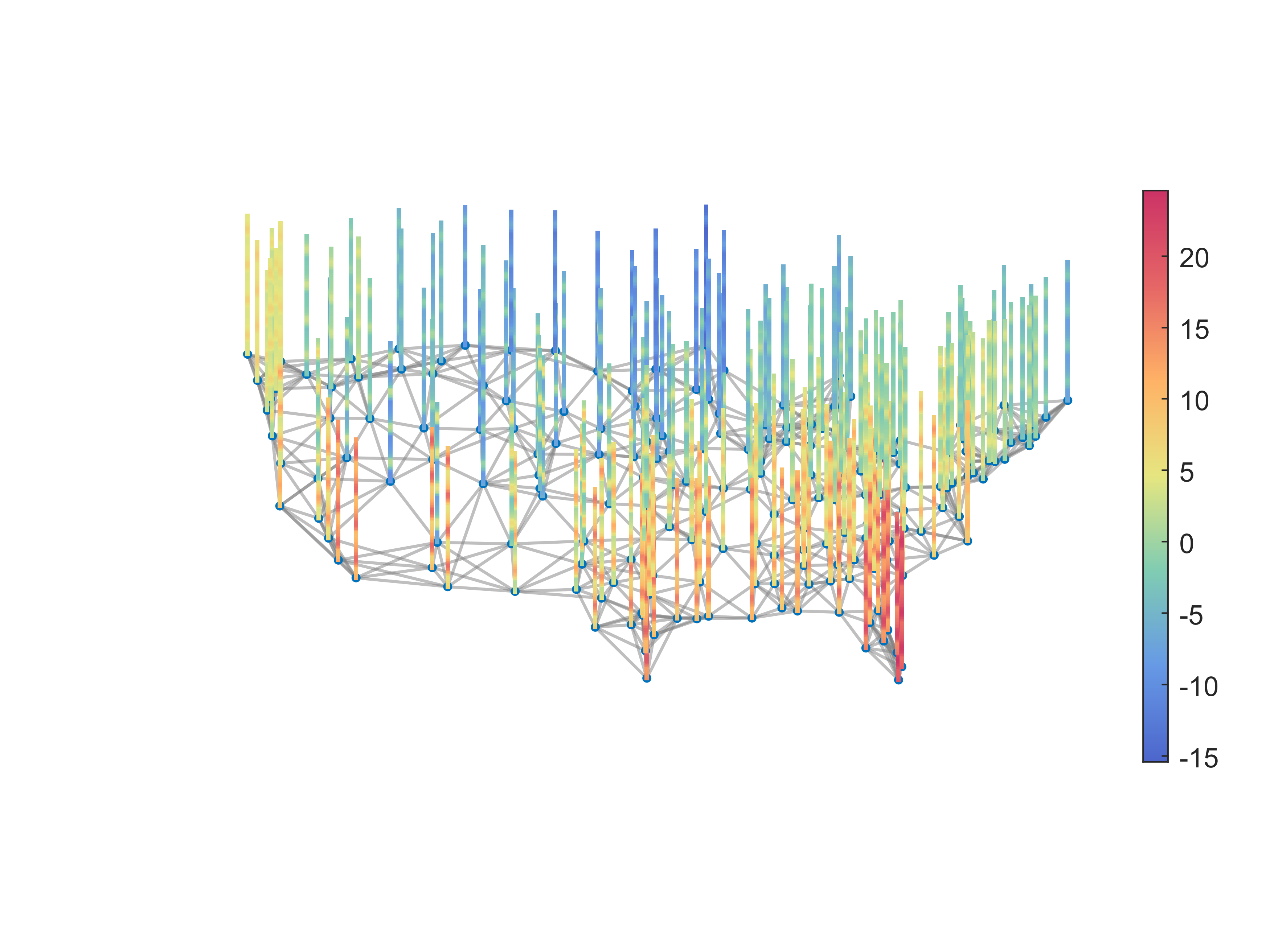}
        \caption{}
        \label{fig_STG_termperature}
    \end{subfigure}
    \hfill
    \begin{subfigure}[t]{0.48\linewidth}
        \centering
        \includegraphics[width=\linewidth]{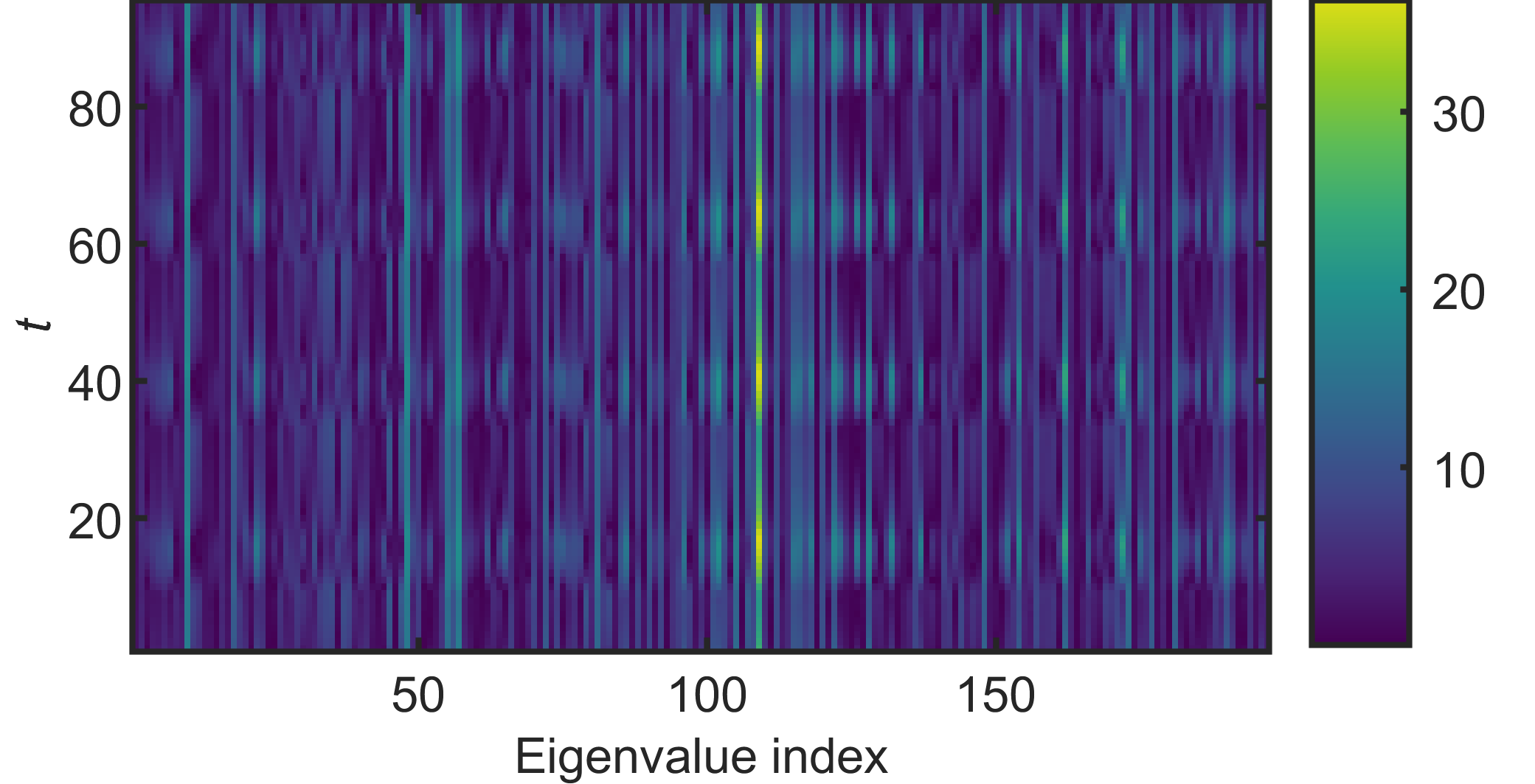}
        \caption{}
        \label{fig_GFT}
    \end{subfigure}
    \caption{Illustrating an STG of hourly temperature. (a) Spatial domain: each vertical bar on a node represents the color coding of the time-varying signal of that node. (b) Spectral-domain: the GFT of the time-varying hourly temperature.}
    \label{fig_combined}
\end{figure*}
Given multi-dimensional data sequence,  $\mathbf{X} = \{\boldsymbol{x}[t] \in \mathbb{R}^{N\times T} | t=1,\cdots, T\}$, where $N$ is the number of vertices and $T$ is the number of time points. 
A TVG is noted by $\mathcal{G}[t] = (\mathcal{V}, \mathcal{E}[t], \mathbf{A}[t])$, where $V$ is the vertex set, $\mathcal{E}[t]$ and $\mathbf{A}[t]$ respectively denote the edge set and adjacency matrix at time t. 
In most TVG modeling conventions, the vertex set $V$ is assumed to remain constant, while the edge set $\mathcal{E}[t]$ may evolve over time.
Accordingly, the associated adjacency matrix $\mathbf{A}[t]$, edge-based operators (e.g., $R_N$), and edge-dependent quantities change consistently with $\mathcal{E}[t]$, whereas node insertion or deletion is typically treated as a separate modeling regime. 

In STGs, the node signals $\boldsymbol{x}[t]$ vary over time, while other graph elements, such as the topology $\mathcal{G}$ or the adjacency matrix $\mathbf{A}$, remain static. 
STGs are also known as the time-vertex graphs, a terminology commonly used in the GSP literature~\cite{Grassi_2018_time_vertex, kartal_2022_joint_time_vertex}, and STGs are more frequently seen in GNN-related research~\cite{liu2019spatiotemporal, jin2023spatio}. 
In GSP, time-varying signals on STGs can be modelled using product graphs. 
Product graphs capture not only correlations between nodes but also correlations across time \cite{Jiang_2021_theory}. 
In both DTDGs and CTDGs, the dynamic behaviors are shown in not only the signals on the graph nodes $\boldsymbol{x}[t]$ but also in the graph topology itself $\mathcal
{G}[t]$. Specifically, in DTDGs, the dynamic graphs are represented by a series of snapshots that can be the graph adjacency matrices and corresponding node feature vectors over time. 
Both STGs and DTDGs are extensively used, as time-varying dynamics are straightforwardly represented as snapshots. By contrast, the CTDGs have emerged in the field of TVGNNs and have yet to be explored in TVGSP. In CTDGs, the TVGs are represented by a more complicated representation in which an initial graph is first defined, and a sequence of graph evolution events defines the time-varying dynamics over time. 

\subsection{Graph Signal Processing and Graph Neural Networks}
The graph Laplacian matrix $\mathbf{L}$, which combines information from $\mathbf{A}$ and $\mathbf{D}$, is defined as $\mathbf{L} = \mathbf{D} - \mathbf{A}$. 
The GFT is based on the eigenvector decomposition 
\begin{equation}
    \mathbf{L} = \mathbf{U}\boldsymbol{\Lambda} \mathbf{U}^\top,
    \label{eq_eigen_decomposition_L}
\end{equation}
where $\mathbf{U}$ is the orthonormal eigenvector matrix and $\boldsymbol{\Lambda}$ is the diagonal matrix of eigenvalues $\Lambda =$ diag$([\lambda_1, \dots, \lambda_N])$.
Notice that for STGs, the GFT in~\eqref{eq_eigen_decomposition_L} is directly applicable to all time points since the Laplacian $\mathbf{L}$ is fixed over time.
For DTDG and CTDG settings, the same definition still holds, but the GFT is applied to each graph realization, meaning $\mathbf{L}[t]$ (and its eigendecomposition) is recalculated whenever the topology changes.

The GFT transforms a graph signal $\boldsymbol{x}$ from the spatial domain to the spectral domain by projecting $\boldsymbol{x}$ onto $\mathbf{U}$: 
\begin{equation}
    \acute{\boldsymbol{x}} = \mathbf{U}^\top \boldsymbol{x}. 
    \label{eq_GFT}
\end{equation}
A graph signal transformed to the spectral domain can be converted back to the spatial domain using the inverse graph Fourier transform (IGFT):
\begin{equation}
    \boldsymbol{x} = \mathbf{U} \acute{\boldsymbol{x}}.
    \label{eq_IGFT}
\end{equation}
The GFT of the TVG signal (hourly temperature) in Fig.~\ref{fig_STG_termperature} is shown in Fig.~\ref{fig_GFT}.
Spectral-domain operations can be performed similarly to those in the classical Fourier transform by defining a filter $\mathbf{H}(\boldsymbol{\Lambda})$, which is then applied using the convolution property of the Fourier transform.
A basic yet complete graph convolution procedure to apply the filter $\mathbf{H}(\boldsymbol{\Lambda})$ to $\boldsymbol{x}$ in the spectral domain and transform it back into the spatial domain:
\begin{equation}
    \boldsymbol{x}_p = \mathbf{U} \mathbf{H}(\boldsymbol{\Lambda}) \mathbf{U}^\top \boldsymbol{x}.
    \label{eq_spectral_graph_convolution}
\end{equation}
In GSP, concepts like low-pass, high-pass, band-pass, and band-reject filters can be analogously implemented through filter design \cite{tremblay_2018_filter_design}.
Wavelets on graphs can be constructed by applying a kernel function to the decomposition of the graph Laplacian $\mathbf{L}$ in \eqref{eq_eigen_decomposition_L} to achieve localized representations \cite{HAMMOND2011_wavelet}.
A graph signal is spectrally sparse when it is bandlimited in the frequency domain. 
For example, a bandlimiting filter $\Sigma$ is defined based on a frequency set $\mathcal{F}$, in which $\mathbf{H}(\lambda) = \Sigma = \text{diag}(1_\mathcal{F}(\lambda))$, with $1_\mathcal{F}(\lambda_i) = 1$ if $\lambda_i \in \mathcal{F}$ and 0 otherwise \cite{Lorenzo_2016_GLMS}. 
A (spectral) GCN layer can be obtained by feeding \eqref{eq_spectral_graph_convolution} into a non-linear activation $\sigma(\cdot)$:
\begin{equation}
    \boldsymbol{x}_{l+1} = \sigma(\mathbf{U H}(\boldsymbol{\Lambda}) \mathbf{U}^\top \boldsymbol{x}_l+\boldsymbol{b}_l),
    \label{eq_spectral_GCN}
\end{equation}
where $\boldsymbol{b}_l$ is an additional bias term to implement, and $\mathbf{ H}(\boldsymbol{\Lambda})$ will now be learned from the back-propagation as learnable parameters instead of being predefined.

The spectral graph convolution can be approximated into a spatial graph convolution by using Chebyshev polynomial approximation for a series of Chebyshev polynomials $T_p(\mathbf{L})$ defined over the interval $[0,\lambda_N]$ \cite{shuman_2011_chebyshev}:
\begin{equation}
    T_p(\mathbf{L}) = 
    \begin{cases}
    1, & \text{if } p = 0\\
    \frac{2\mathbf{L}-\lambda_N}{\lambda_N}, & \text{if } p = 1\\
    \frac{4\mathbf{L}-2\lambda_N}{\lambda_N}T_{p-1}(\mathbf{L})-T_{p-2}(\mathbf{L}), & \text{if } p \geq 2.
    \end{cases}
\end{equation}
with the approximation being
\begin{equation}
    \mathbf{U}\mathbf{H}(\boldsymbol{\Lambda})\mathbf{U}^\top = \mathbf{H}(\mathbf{L}) \approx \theta_0+\sum_{p=1}^{P}\theta_p T_p(\mathbf{L}), 
    \label{eq_cheb_H_1}
\end{equation}
where $P$ is the number of polynomials, $\theta_p$ and $\hat{\theta}_p$ are the weight of the $p^\text{th}$ polynomial. 

A graph signal with a reduced number of nodes, sampled or masked according to a node sampling set $S \subseteq V$, is sparse in the spatial domain.
The matrix $\mathbf{D}_S$ represents the sampling or masking matrix, containing only non-zero diagonal elements given by $\mathbf{D}_{S_{ii}} = 1 \forall v_i \in S$. 
The matrix $\mathbf{D}_S$ can be designed to introduce spatial sparsity based on a sampling strategy, or it can be used as a masking matrix to record the observation state of the nodes. 
Similar to how the (spectral) GCN is defined in \eqref{eq_spectral_GCN}, a (spatial) GCN can be defined using \eqref{eq_cheb_H_1}:
\begin{equation}
\boldsymbol{x}_{l+1}=\sigma\left(\sum_{p=1}^{P}\theta_p \mathbf{L}^p+\boldsymbol{b}\right), 
    \label{eq_spatial_GCN}
\end{equation}
where parameters $\theta_p$ are the learnable GCN parameters and $\boldsymbol{b}$ is again the bias term.
It should be emphasized that equations \eqref{eq_spectral_GCN} and \eqref{eq_spatial_GCN} establish the basic but direct relationship between GSP and GNNs, making GNNs interpretable using GSP techniques.
GSP interpretations of more advanced GNNs can be achieved through appropriate redefinition of trainable parameters, modification of core components, and integration of machine learning techniques from \eqref{eq_spectral_GCN} and \eqref{eq_spatial_GCN}.

What differentiates GSP from conventional signal processing or standard multivariate algorithms is its integration of graph structures and topological relationships alongside signal data. 
This distinct approach leverages the underlying graph topology to enhance data representation and analysis in GSP \cite{mateos2019connecting}; thus, methods that use conventional spectral transformations, such as DFT, will be omitted.



\section{Signal Processing for Time-varying Graphs}
\label{sec_GSP}
When dealing with TVG signals, GSP algorithms can be broadly categorized into online and offline methods. 
Online TVGSP algorithms, such as adaptive filters and state-space models, update dynamically with time-varying data, offering real-time processing capabilities. 
Offline methods, on the other hand, analyze graph signals in batches, leveraging the entire dataset to capture spatiotemporal relationships. 
In this section, we will focus on non-GNN approaches, covering algorithm genres such as adaptive filtering, time-series models, and their extensions for graph-structured data.

\begin{figure*}[htb]
    \centering
    \includegraphics[trim={15 390 25 70},clip, width=\linewidth]{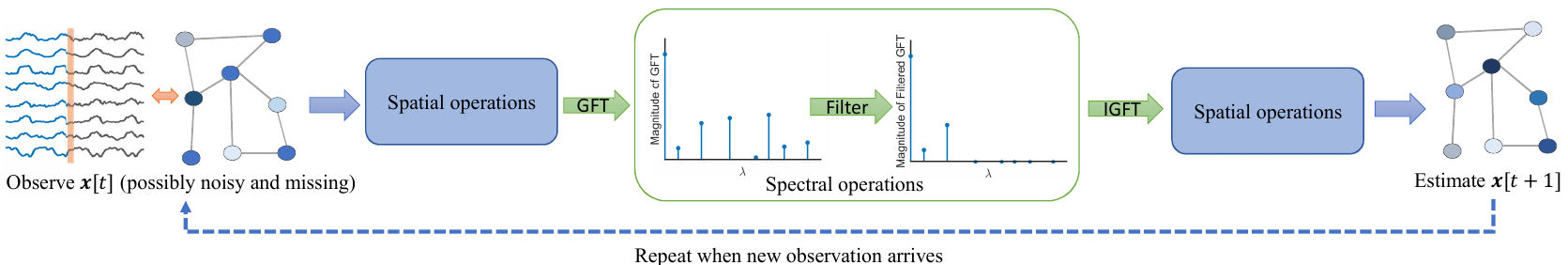}
    \caption{Workflow of online TVGSP algorithms for processing TVG signals on an STG.}
    \label{fig_online_GSP_flow}
\end{figure*}

\subsection{Online TVGSP Algorithms}
Let us begin with the discussion of online TVGSP algorithms for TVG signals.
The online TVGSP algorithms share a workflow similar to that of the online algorithms found in non-graph contexts.
Following the convention seen in most adaptive GSP algorithms, we assume that the data sequence $\mathbf{X} = \{\boldsymbol{x}[t] \in \mathbb{R}^{N\times T} | t=1,\cdots, T\}$ arrives one instance at a time, with each instance will be denoted as $\boldsymbol{x}[t]$.
In some cases, the data can be partially observed, in which the partial observation can be modeled by applying a sampling matrix $\mathbf{D}_S$ to the received time instance $\boldsymbol{x}[t]$.
The data received can also contain noise, which is often represented as an additive noise term $\boldsymbol{\xi}[t]$.
Typically, there will be a weight term defined by utilizing the graph $\mathcal{G}$, which serves as how the observation is modified or processed.
Without the loss of generality, we denote this weight term as $\mathbf{W}$, but keep in mind that the type and dimension of this term are algorithm-specific. 
In many cases, the weight is updated based on calculating a metric between the estimation $\hat{\boldsymbol{x}}[t]$ and the observation ${\boldsymbol{x}}[t]$.
The online TVGSP algorithm update under noisy and missing data observations can be generalized using the following framework:
\begin{equation}
\begin{aligned}
    \hat{\boldsymbol{x}}[t+1] &= f\left(\mathbf{W}[t], \mathbf{D}_S(\boldsymbol{x}[t] + \boldsymbol{\xi}[t])\right) \text{ with} \\
    \mathbf{W}[t] &= g\left(\hat{\boldsymbol{x}}[t], \mathbf{D}_S(\boldsymbol{x}[t] + \boldsymbol{\xi}[t])\right),
\end{aligned}
\label{eq_online_GSP_framework}
\end{equation}
where $f(\cdot)$ denotes the graph signal update function and $g(\cdot)$ denotes the weight update function. 
An illustration of the workflow of online TVGSP algorithms is shown in Fig.~\ref{fig_online_GSP_flow}.

\subsubsection{Spectral Adaptive Filters} A natural idea for the online processing of time-varying node signals is to utilize adaptive graph filters that consist of a combination of classical adaptive filters with graph shift operations and can efficiently process time-varying node signals owing to their simplicity of implementation. 
Similar to the classical adaptive filters, GSP-based adaptive filters update in the direction opposite to the error at each time step based on convex optimization. 
What makes this combination different from the classical adaptive filters is that most GSP-based adaptive filters utilize a predefined bandlimited filter in the spectral domain derived from GFT, which implies that it captures the topological information of the graph in the spectral domain along with the time-varying signal. 
Using the framework we defined in \eqref{eq_online_GSP_framework} and the graph convolution in \eqref{eq_spectral_graph_convolution}, the weight in GSP-based adaptive filters on STGs can be expressed as $\mathbf{W}[t] = \mathbf{U H}(\boldsymbol{\Lambda}, t) \mathbf{U}^\top$.
By leveraging topological information, GSP-based adaptive filters enable signal reconstruction across both spatial and temporal dimensions.
The first GSP-based adaptive filter is the adaptive graph Least Mean Squares (GLMS) algorithm\cite{Lorenzo_2016_GLMS}.
In the GLMS, the update strategy is the optimization results derived based on an $l2$-norm minimization with the assumption of Gaussian noise:
\begin{equation}
    \mathcal{L}(\hat{\boldsymbol{x}}[t])=\frac{1}{2} \mathbb{E}\left\Vert \mathbf{D}_S(\boldsymbol{x}[t] + \boldsymbol{\xi}[t])-\mathbf{D}_S\mathbf{W}\hat{\boldsymbol{x}}[t]\right\Vert_2^2, \label{eq_lms_cost} 
\end{equation}
where $ \mathbb{E}$ is the expectation, $\mathbf{W} = \mathbf{U H}(\boldsymbol{\Lambda}) \mathbf{U}^\top$ and $\mathbf{H}(\boldsymbol{\Lambda})$ is the static predefined bandlimited filter. 
As a general rule of thumb, the update function $f(\cdot)$ of the GSP-based adaptive filter can be derived using gradient methods:
\begin{equation}
    \hat{\boldsymbol{x}}\left[t+1\right]= f\left(\mathbf{W}, \mathbf{D}_S(\boldsymbol{x}[t] + \boldsymbol{\xi}[t])\right)
    = \hat{\boldsymbol{x}}\left[t\right]-\mu\frac{\partial \mathcal{L}(\hat{\boldsymbol{x}}[t])}{\partial \hat{\boldsymbol{x}}[t]}
    \label{eq_adaptive_GSP_gradient}
\end{equation}
The plugging \eqref{eq_lms_cost} into \eqref{eq_adaptive_GSP_gradient}, update function of GLMS  is
\begin{equation}
    f\left(\mathbf{W}, \mathbf{D}_S(\boldsymbol{x}[t] + \boldsymbol{\xi}[t])\right) = \hat{\boldsymbol{x}}[t]+\mu\mathbf{WD}_S
    \left(\boldsymbol{y}[t]-\hat{\boldsymbol{x}}[t]\right),
    \label{eq_glms_update}
\end{equation}
where $\mu$ is the step-size parameter.
There are several variants and extensions of the GLMS. 
By introducing small perturbations to the Laplacian matrix, the GLMS can be extended to dynamic graphs \cite{Lorenzo_2018_online_dynamic_LMS}. 
A variant of GLMS is  Graph Normalized LMS (GNLMS), in which a spectral domain normalization is derived from minimizing the spectral error between consecutive estimations and then introduced to the weights $\mathbf{W}$ of the LMS-based update function to improve the convergence behavior \cite{Spelta_2020_NLMS}.
Graph Recursive Least Squares extends the RLS framework to enhance convergence and prediction under noisy, time-varying signals by recursively minimizing a weighted error criterion, outperforming GLMS \cite{Lorenzo_2017_RLS, Sadigh_GRLS_2024}.

The Gaussian noise assumption is seen in most noise models, and $l_2$-norm minimization is the go-to option for Gaussian noise because minimizing the squared error corresponds to the maximum likelihood estimate solution \cite{Chen_2016_variance}.
However, the underlying noise in a variety of realistic applications, including meteorological recordings \cite{1986_weather_impulsive} and powerline communication \cite{karakucs_2020_modelling}, is verified to possess impulsive behaviors that could be represented by heavy-tailed, non-Gaussian distributions, 
such as generalized Gaussian, Student's t, and $\alpha$-stable distributions \cite{Chen_2016_variance}. 
The least-squares-based methods lose their validity in impulsive noise situations owing to the presence of large or infinite variance caused by the impulsiveness in the noise \cite{kuruouglu_1998_least}. 
Several approaches can be taken to overcome this drawback.
Instead of using the second power, the least mean $p^\text{th}$ power is used to define a loss based on the assumption that the noise is a symmetric $\alpha$-stable distribution (S$\alpha$S), giving us the Graph Least Mean $p^\text{th}$ algorithm \cite{nguyen2020_LMP}: 
\begin{equation}
    \mathcal{L}(\hat{\boldsymbol{x}}[t])=\frac{1}{p} \mathbb{E}\left\Vert \mathbf{D}_S(\boldsymbol{x}[t] + \boldsymbol{\xi}[t])-\mathbf{D}_S\mathbf{W}\hat{\boldsymbol{x}}[t]\right\Vert_p^p + r(\hat{\boldsymbol{x}}[t]), \label{eq_lmp_cost} 
\end{equation}
with $1 \leq p \leq 2 $ and $r(\hat{\boldsymbol{x}}[t])$ being the regularization term.
The GLMP update function following the framework shown in \eqref{eq_online_GSP_framework} can be obtained by plugging \eqref{eq_lmp_cost} into \eqref{eq_adaptive_GSP_gradient}: 
\begin{equation}
\begin{split}
            &f\left(\mathbf{W}, \mathbf{D}_S(\boldsymbol{x}[t] + \boldsymbol{\xi}[t])\right) = \hat{\boldsymbol{x}}[t] \\ 
            &+\mu\mathbf{W}
    \text{sign}(\mathbf{D}_S(\boldsymbol{y}[t]-\hat{\boldsymbol{x}}[t])) \circ |\mathbf{D}_S(\boldsymbol{y}[t]-\hat{\boldsymbol{x}}[t])|^{p-1},
\end{split}
    \label{eq_glmp_update}
\end{equation}
where $\circ$ is the Hadamard product and the superscript $^{p-1}$ denotes element wise $p-1^\text{th}$ power.
The Normalized GLMP was further proposed to include a time-varying spectral normalization matrix to speed up the computation of GLMP  \cite{yan_2022_NLMP}.  
Moving further with the idea of $l_p$-norm minimization, the adaptive Graph-Sign algorithm (Graph-Sign) was derived as a graph extension of the classical adaptive sign error or the least mean absolute deviation (LMAD) algorithm for multivariate signals \cite{yan_2022_sign, bib_MD_LMAD}. 
The Graph-Sign algorithm is derived from the minimum dispersion criterion and then reduced to $l_1$-norm minimization, which removes the need for prior knowledge from any noise assumption \cite{bib_l1}. 
The resulting update algorithm is 
\eqref{eq_online_GSP_framework} can be obtained by plugging \eqref{eq_lmp_cost} into \eqref{eq_adaptive_GSP_gradient}: 
\begin{equation}
        f\left(\mathbf{W}, \mathbf{D}_S(\boldsymbol{x}[t] + \boldsymbol{\xi}[t])\right) = \hat{\boldsymbol{x}}[t]+\mu\mathbf{W}
    \text{sign}(\mathbf{D}_S(\boldsymbol{y}[t]-\hat{\boldsymbol{x}}[t])).
    \label{eq_gsign_update}
\end{equation}
Because the Graph-Sign is obtained from an $l_1$-norm problem, the resulting update is robust.
To view it from another angle, the Sign$(\cdot)$ function is a clipping function, which is not affected by the outliers and extreme values in the error $\boldsymbol{e}[t]$, making the Graph-Sign robust. 
This allows the Graph-Sign algorithm to avoid the instability seen in the least-squares-based algorithms when estimating graph signals under impulsive noise. 

An alternative approach to simply modifying the loss function by switching $l_2$-norm to $l_1$-norm or $l_p$-norm to enhance robustness is to employ a robust adaptive estimation of graph signals developed by replacing the convention of using $l_2$-norm to define the optimization problem with the Welsch loss \cite{wang2022robust}. 
The Welsch loss differs from conventional loss functions by incorporating robustness as a continuous parameter, allowing the Welsch loss to adapt dynamically during algorithm deployment \cite{barron_2019_general}. 
Instead of directly changing the cost function, a less extreme method to the problem is to add a regularization term to the LMS cost function. 
The Graph Alternating Minimization algorithm improves the robustness of GLMS by separately modeling the impact of Gaussian noise and outliers in the cost function \cite{Li_robust_2023}. 
The graph median filter has also shown good performance under impulse noise \cite{tay2020time, tay_2021_sensor}.


\textbf{Extending adaptive filters to DTDGs and CTDGs}: The adaptive GSP filters reviewed here were initially proposed for STGs; however, their formulation is readily generalizable to DTDGs by introducing time dependency into the graph operators and filter coefficients.
At present, however, to the best of our knowledge, this generalization has not been systematically formalized in the literature, and existing work rarely addresses adaptive filtering explicitly in dynamic graph settings. 
Our discussion here highlights an important and largely unexplored research direction, where the adaptive TVGSP framework could be extended to support time-evolving graphs.
 
Given a DTDG with graph $\mathcal{G}[t]$ at time $t$, to define the GSP-based adaptive filters, we can replace $\mathbf{W} = \mathbf{U H}(\boldsymbol{\Lambda}) \mathbf{U}^\top$ in \eqref{eq_adaptive_GSP_gradient} with 
\begin{equation}
    \mathbf{W}[t] = \mathbf{U}[t] \mathbf{H}[t](\boldsymbol{\Lambda})^\top\mathbf{U}[t]^\top,
    \label{eq_GSP_DTDG_spectral}
\end{equation}
where $\mathbf{U}[t]$ is obtained from the GFT of $\mathcal{G}[t]$, which is from the eigenvector decomposition of the instantaneous graph Laplacian matrix $\mathbf{L}[t]$.

Alternatively, instead of defining the spectral operator directly from the instantaneous Laplacian at each time, we can view the DTDG as producing incremental changes with respect to the previous graph. Let
\begin{equation}
\mathbf{W}[t-1] = \mathbf{U}[t-1]\mathbf{H}[t-1](\boldsymbol{\Lambda}[t-1])\mathbf{U}[t-1]^\top
\end{equation}
denote the adaptive GSP filter at time $t-1$. When the graph evolves to $\mathcal{G}[t]$, we write
\begin{equation}
\mathbf{W}[t] = \mathbf{W}[t-1] + \Delta_{\mathbf{W}}[t],
\end{equation}
where $\Delta_{\mathbf{W}}[t]$ captures the effect of the topology change, for example via a change of the GFT basis,
\begin{equation}
\mathbf{U}[t] = \mathbf{U}[t-1] + \Delta_{\mathbf{U}}[t], 
\end{equation}
and the change of the operator
\begin{equation}
    \mathbf{H}[t](\boldsymbol{\Lambda}[t]) 
= \mathbf{H}[t-1](\boldsymbol{\Lambda}[t-1]) + \Delta_{\mathbf{H}}[t].
\label{eq_GSP_DTDG_filter}
\end{equation}

To provide a demonstration using GLMS in \eqref{eq_glms_update}, the DTDG version can be expressed as  
\begin{equation}
    f\left(\mathbf{W}[t], \mathbf{D}_S(\boldsymbol{x}[t] + \boldsymbol{\xi}[t])\right) = \hat{\boldsymbol{x}}[t]+\mu\mathbf{W}[t]\mathbf{D}_S
    \left(\boldsymbol{y}[t]-\hat{\boldsymbol{x}}[t]\right). 
    \label{eq_glms_update_TV}
\end{equation}
Similarly, the DTDG version of GLMP is
\begin{equation}
\begin{split}
    &f\left(\mathbf{W}[t], \mathbf{D}_S(\boldsymbol{x}[t] + \boldsymbol{\xi}[t])\right) = \hat{\boldsymbol{x}}[t] \\&+\mu\mathbf{W}[t]
    \text{sign}(\mathbf{D}_S(\boldsymbol{y}[t]-\hat{\boldsymbol{x}}[t])) \circ |\mathbf{D}_S(\boldsymbol{y}[t]-\hat{\boldsymbol{x}}[t])|^{p-1},
\end{split}
\label{eq_glmp_update_TV}
\end{equation}
and the DTDG version of Graph-Sign is 
\begin{equation}
\begin{split}
     f\left(\mathbf{W}[t], \mathbf{D}_S(\boldsymbol{x}[t] + \boldsymbol{\xi}[t])\right) 
    = \hat{\boldsymbol{x}}[t] +\mu\mathbf{W}[t] \text{sign}(\mathbf{D}_S(\boldsymbol{y}[t]-\hat{\boldsymbol{x}}[t])).
\end{split}
\label{eq_gsign_update_TV}
\end{equation}

For the CTDG counterpart, it is more natural to model changes as event-driven 
rather than discrete temporal updates. 
In this case, instead of 
indexing the adaptive GSP filter by discrete time instants, we associate each 
update with an event $\mathcal{E}_k$ that modifies the graph topology. 
Let
\begin{equation}
    \mathbf{W}[\mathcal{E}_{k-1}] 
    = \mathbf{U}[\mathcal{E}_{k-1}]\,\mathbf{H}[\mathcal{E}_{k-1}](\boldsymbol{\Lambda}[\mathcal{E}_{k-1}])\,
    \mathbf{U}[\mathcal{E}_{k-1}]^\top,
    \label{eq_GSP_CTDG_spectral}
\end{equation}
denote the adaptive spectral filter immediately before the event 
$\mathcal{E}_k$ occurs. 
When the graph evolves due to event $\mathcal{E}_k$, the updated filter can be written as
\begin{equation}
    \mathbf{W}[\mathcal{E}_k] 
    = \mathbf{W}[\mathcal{E}_{k-1}] + \Delta_\mathbf{W}[\mathcal{E}_k],
\end{equation}
where $\Delta_\mathbf{W}[\mathcal{E}_k]$ captures the spectral effect of the event-driven topology change. This perturbation may arise from a change in the 
GFT basis,
\begin{equation}
    \mathbf{U}[\mathcal{E}_k] 
    = \mathbf{U}[\mathcal{E}_{k-1}] + \Delta_\mathbf{U}[\mathcal{E}_k],
\end{equation}
and a corresponding update of the spectral operator,
\begin{equation}
\mathbf{H}[\mathcal{E}_k](\boldsymbol{\Lambda}[\mathcal{E}_k])
=
\mathbf{H}[\mathcal{E}_{k-1}](\boldsymbol{\Lambda}[\mathcal{E}_{k-1}])
+ \Delta_\mathbf{H}[\mathcal{E}_k].
\label{eq_GSP_CTDG_filter}
\end{equation}
This event-based formulation highlights how CTDG evolution can be incorporated 
into adaptive GSP filtering by explicitly linking filter updates to discrete 
structural events rather than uniform temporal sampling.

Compared with the STG case, extending adaptive spectral GSP methods to DTDGs and CTDGs requires recomputing the GFT whenever the topology of $\mathcal{G}[t]$ changes, resulting in a computational burden on the order of $O(N^3)$ over each evolving graph. This becomes challenging for large-scale or rapidly evolving systems, and repeated eigendecompositions may also introduce numerical instability for large $N$. 
Currently, extensions of TVGSP to DTDGs and CTDGs remain limited due to this scalability issue, with existing approaches such as Sardellitti et al. ~\cite{sardellitti_2021_online_small_pertubation} where topology changes through small perturbations that allow only 1 edge update per topology change.
On the other hand, these challenges highlight a compelling opportunity for developing more stable, efficient, and theoretically grounded graph operators tailored to dynamic graph settings.
Designing such scalable GFT surrogates and adaptive spectral frameworks for TVGs remains an exciting and largely open research direction.

\subsubsection{Spatial Adaptive Filters} Notice that the adaptive GSP algorithms are not limited to the spectral domain but could also be conducted in the spatial domain by using the Chebyshev polynomial approximation shown in \eqref{eq_cheb_H_1}. 
For example, the bandlimited filters in the GLMS can be approximated using a series of Chebyshev polynomials, leading to the spatial domain graph diffusion \cite{Roula_2017_LMS_Diffusion}. 
A GSP-based adaptive filter that uses a combination of graph diffusion and sign update strategy was recently proposed in \cite{yan_2023_sign}. 
Recently, the Graph Signal Adaptive Message Passing has been proposed using the flexible localized node message passing scheme instead of defining all computations globally on the graph level \cite{yan_2025_GSAMP}. 
In general, we can use the Chebyshev polynomial approximation shown in \eqref{eq_cheb_H_1} to define the weight in the cost function, such as the ones seen in \eqref{eq_lms_cost} and \eqref{eq_lmp_cost}:
\begin{equation}
    \mathbf{W} \approx\theta_0+\sum_{p=1}^{P}\theta_p T_p(\mathbf{L}) = \sum_{p=0}^{P}\hat{\theta}_p \mathbf{L}^p,
    \label{eq_adaptive_spatial_weight}
\end{equation}
where $\hat{\theta}_p$ are the coefficients for $\mathbf{L}^p$ after merging the terms within $T_p(\mathbf{L})$. 
It is worth mentioning that the spatial methods resemble the structure of (spatial) GCN-based GNNs: a GCN layer is obtained by algebraic manipulations of \eqref{eq_cheb_H_1}, then feeding its output into an activation function \cite{kipf2017semi_supervised, defferrard2016_cheb}. 
We will conduct a more thorough review of GNN-based methods for TVG signals in Section~\ref{section_GNN}. 

We can again define the DTDG versions of the GSP-based spatial adaptive filters using logic similar to how the DTDG GSP-based (spectral) adaptive filters are defined.
In the case of the spatial algorithms, for a DTDG $\mathcal
{G}[t]$, all the operations are directly defined by the graph Laplacian matrix $\mathbf{L}[t]$, meaning that we replace $\mathbf{W}$ in \eqref{eq_adaptive_spatial_weight} with 
\begin{equation}
    \mathbf{W}[t] \approx\theta_0[t]+\sum_{p=1}^{P}\theta_p[t] T_p(\mathbf{L}[t]) = \sum_{p=0}^{P}\hat{\theta}_p[t] \mathbf{L}^p[t].
    \label{eq_GSP_DTDG_spatial}
\end{equation}

Alternatively, instead of constructing the spatial operator directly from the
instantaneous Laplacian $\mathbf{L}[t]$, one may interpret the evolution of the
DTDG as producing incremental changes with respect to the operator at the
previous time step. Let
\begin{equation}
\mathbf{W}[t-1]
=
\sum_{p=0}^{P}\hat{\theta}_p[t-1]\,
\mathbf{L}^{p}[t-1]
\end{equation}
denote the spatial GSP adaptive filter at time $t-1$. When the graph evolves to
$\mathcal{G}[t]$, we write
\begin{equation}
\mathbf{W}[t]=\mathbf{W}[t-1]+\Delta_{\mathbf{W}}[t],
\end{equation}
where $\Delta_{\mathbf{W}}[t]$ accounts for the effect of the topology variation.
For instance, this can be modeled via a temporal-modification of the Laplacian,
\begin{equation}
\mathbf{L}[t]
=
\mathbf{L}[t-1]
+
\Delta_\mathbf{L}[t],
\end{equation}
and the corresponding variation of the filter coefficients
\begin{equation}
\hat{\theta}_p[t]
=
\hat{\theta}_p[t-1]
+
\Delta\hat{\theta}_p[t], \qquad p = 0,\dots,P.
\end{equation}

Similar to the spectral case, the CTDG version of spatial adaptive filters has not yet been fully realized, largely due to the difficulty of event-driven graph updates within a stable GSP representation framework. 
Recall that spatial filtering can be related to spectral filtering through the Chebyshev polynomial approximation in~\eqref{eq_cheb_H_1}, which effectively serves as a bridge between spectral operators and spatial-domain implementations. 
Building on this connection, one could follow a similar line of reasoning as in the DTDG case: an event-driven formulation could then be derived from~\eqref{eq_GSP_CTDG_spectral} to a corresponding spatial adaptive operator in~\eqref{eq_GSP_CTDG_filter}. 

\subsubsection{State Space Models for TVGSP} Aside from GSP-based adaptive filters, another direction to approach the problem is to use state space models, and the typical example is the Kalman Filter \cite{kalman_1960_new}. 
The Kalman Filter is a recursive algorithm used to estimate the state of a linear dynamic system under the Gaussian noise assumption.  
In GSP, we can formulate system equations suitable for TVG signals by slightly modifying the framework in \eqref{eq_online_GSP_framework}:
\begin{equation}
    \begin{aligned}
    \boldsymbol{x}[t] &= f_{\text{state}}(\mathbf{L}, \boldsymbol{x}[t-1]) + \boldsymbol{\xi}_{\text{state}}[t] \text{ and} \\
    \boldsymbol{y}[t] &= g_{\text{obs}}(\mathbf{L}, \boldsymbol{x}[t]) + \boldsymbol{\xi}_{\text{obs}}[t], 
    \end{aligned}
    \label{eq_kalman_system}
\end{equation}
where $f_{\text{state}}(\mathbf{L}, \boldsymbol{x}[t-1])$ represents the state evolution function, and $g_{\text{obs}}(\mathbf{L}, \boldsymbol{x}[t])$ represents the observation function. Both functions operate on graph signals using the Laplacian matrix $\mathbf{L}$. 
In the system equations \eqref{eq_kalman_system}, the terms $\boldsymbol{\xi}_{\text{state}}[t]$ and $\boldsymbol{\xi}_{\text{obs}}[t]$ are independent white Gaussian noises with covariance matrices $\mathbf{C}_{\text{state}}[t]$ and $\mathbf{C}_{\text{obs}}[t]$, respectively.

Following the approach commonly used in GSP-adaptive filters, the graph structure can be fully leveraged by applying the GFT in \eqref{eq_GFT}, which transforms the system equations in \eqref{eq_kalman_system} into the spectral domain \cite{Sagi_2023_graph_EFK_Kalman}:
\begin{equation}
    \begin{aligned}
    \acute{\boldsymbol{x}}[t] &= \acute{f}_{\text{state}}(\mathbf{L}, \boldsymbol{x}[t-1]) + \acute{\boldsymbol{\xi}}_{\text{state}}[t] \text{ and} \\
    \acute{\boldsymbol{y}}[t] &= \acute{g}_{\text{obs}}(\mathbf{L}, \boldsymbol{x}[t]) + \acute{\boldsymbol{\xi}}_{\text{obs}}[t]. 
    \end{aligned}
    \label{eq_kalman_system_spectral}
\end{equation}
The operations within $f_{\text{state}}(\mathbf{L}, \boldsymbol{x}[t-1])$ and $g_{\text{obs}}(\mathbf{L}, \boldsymbol{x}[t])$ can now be conducted entirely in the spectral domain using spectral graph convolution in  \eqref{eq_spectral_graph_convolution}. 
With the multivariate signals represented as graph signals, the rest of the procedure, such as the update step and the prediction step, follows the classical Kalman filter workflow, with adjustments for the graph representation. 
Once the estimation and update steps are completed, the results are transformed back into the spatial domain using the IGFT in \eqref{eq_IGFT}.  

The Kalman Filter over signals on sensor networks can be established, for which the optimal estimation method at each sensor is done when the sensor communications are formed by an undirected graph \cite{Shi_2009_Kalman_filtering_over_graphs}. 
The simplicial complex version of Kalman filters on graphs is also realized through the Hodge Laplacian representation \cite{Money_simplicial_kalman_2023}.
The Extended Kalman filter (EKF) extends the applicability of the standard Kalman filter from linear systems to non-linear systems by locally approximating them as linear \cite{ribeiro2004kalman}.
The graph version of the Extended Kalman filter is a graph process obeying a nonlinear state-space model that can be found by forming a GSP-based Bayesian filter whose goal is to minimize the computed covariance matrix as the Kalman gain of the Extended Kalman filter in the graph spectral domain \cite{Sagi_2023_graph_EFK_Kalman}. Recent work further extends EKF to incorporate graph topology information via the graph Laplacian, enabling more robust localization in networked dynamic systems. For instance, in the context of connected and automated vehicles, the graph Laplacian Extended Kalman Filter leverages the structural relations between vehicles to enhance estimation accuracy and stability under nonlinear dynamics \cite{piperigkos2021graph}. 
The Unscented Kalman Filter (UKF) addresses the non-linearity of systems by representing the mean and covariance of the state using a set of deterministically chosen sample points to avoid the need for linearization, which is different from the standard Kalman filter \cite{wan_2001_unscented}. 
This approach can also be applied to graph signals, which leads to the Unscented Kalman filter of graph signals by again using the decomposition of the graph Laplacian matrix to form the GFT and then designing the Kalman gain matrix from the eigenvectors \cite{li2023unscented}. 
The Kalman filter is designed for Gaussian noise because of minimizes the mean square error of the estimated states, which is not suitable for non-Gaussian noise, making this research challenge to be addressed in the future.   

Beyond the Kalman filtering framework, particle filters provide a more general approach capable of handling non-Gaussian and highly nonlinear systems. When extended to graph settings, graph particle filters utilize the underlying topology to improve estimation. In particular, the graph Laplacian distributed particle filter promotes smoothness of state estimates across the network by regularizing neighboring particle estimates based on the graph structure \cite{rabbat2016graph}. To further address communication efficiency in distributed implementations, graph-based compression schemes have been proposed. These methods transform local particle distributions into the graph frequency domain to reduce communication overhead while preserving estimation accuracy \cite{yu2019graph}. Together, these advancements illustrate how graph structure can enhance both Kalman and particle filtering methods, offering promising directions for state estimation in complex networked systems.

Similar to GSP-based adaptive filters, the DTDG and CTDG versions of state-space models for TVGSP are limited. 
Here, we outline a general approach to realizing state-space models for DTDGs. 
Topology-based parameters such as $\mathbf{L}$ can be replaced with their time-varying counterparts $\mathbf{L}[t]$ to account for the topology changes induced to incorporate the topology change caused by $\mathcal{G}$. 
However, we will not proceed with detailed derivations, proofs, or justifications at this stage, as merging topological changes with the inherently dynamic parameter settings of state-space models poses significant challenges and is left for future research.
For CTDGs, the additional complexity of integrating events with GSP and state-space models presents an even greater challenge, requiring dedicated exploration for future research.

\textbf{Remarks} Online algorithms are most preferred when the outputs are required to be made on the fly as new inputs are received. 
Their ability to produce real-time outputs based on very limited data observations in the temporal domain, without requiring the entire input upfront, makes them highly effective in TVGSP tasks due to the fact that most TVG signals already have a large spatial dimension.
Online TVGSP algorithms are not limited to regression tasks; their outputs can also serve as inputs for classification or clustering models \cite{saboksayr_2021_online_discriminative_graph_learning, Dong_Graph_ML_2020}. 
This capability broadens the application scope of online TVGSP algorithms. 
On one hand, online algorithms often have lower computational complexity to cope with real-time output capability, which is a critical advantage of deploying online algorithms on TVG signals. 
On the other hand, this same online requirement often restricts the length of the prediction horizon of online algorithms, limiting their ability to generate long-term temporal predictions. 
In contrast, offline algorithms process the entire dataset at once, enabling long-term predictions on graph signals.
In the next subsection, we will be discussing offline GSP-based algorithms for TVG signals.

\begin{figure*}[htb]
    \centering
    \includegraphics[trim={15 400 70 70},clip, width=\linewidth]{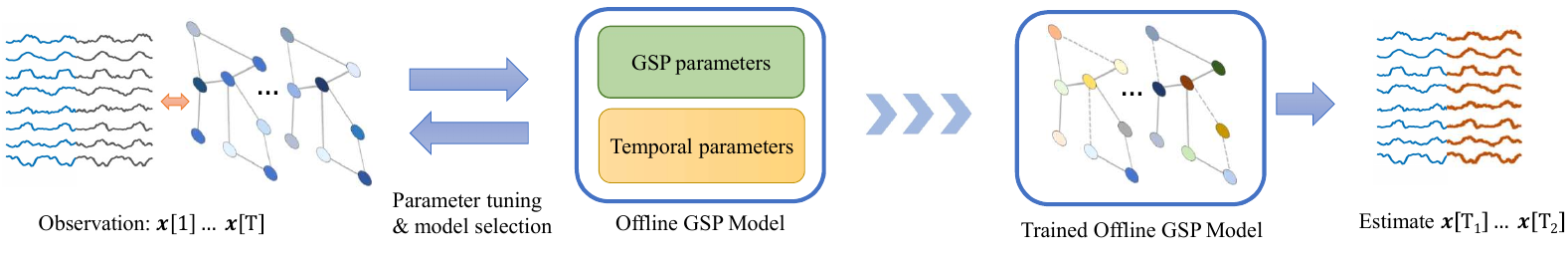}
    \caption{Workflow of offline TVGSP algorithms for processing TVG signals on an STG.}
    \label{fig_offline_GSP_flow}
\end{figure*}

\subsection{Offline GSP Algorithms}

\subsubsection{General Offline GSP}
A naive way to process the time-varying signals is to apply the graph convolution in \eqref{eq_spectral_graph_convolution} or a GCN to the observations as we receive new observations of the graph signal. 
However, this method does not capture any temporal changes in the signal and often results in poor performance. 
It is essential to capture both spatial and temporal dynamics to effectively process time-varying signals on graphs.

Another way to use GSP to process TVG signals is to treat each time instance as a feature and formulate a multi-feature GSP cost function to minimize the error between estimation $\hat{\mathbf{X}}$ and ground truth $\mathbf{X}$: 
\begin{equation}
    J(\hat{\mathbf{X}}) = \|\hat{\mathbf{X}} - \mathbf{X}\|_F^2 + \mu \text{Tr} (\hat{\mathbf{X}}^\top \mathbf{L} \hat{\mathbf{X}}),
    \label{eq_offline_cost_bad}
\end{equation}
where $\|\cdot\|^2_F$ is the Frobenius Norm, $\mu$ is the regularization weight, \text{Tr}$(\cdot)$ is the trace, and $\text{Tr} (\hat{\mathbf{X}}^\top \mathbf{L} \hat{\mathbf{X}})$ is a spatial-smoothness-based regularization that minimizes the total variation of the results \cite{Ortega_graph_2018}.
The solution can be found by calculating the gradient and setting it to zero. 
This method has 2 potential drawbacks.
First, treating signals from different time instances as features means that the method is still inherently time-invariant, which does not take into account any temporal dynamics within the signal.
Second, after some algebraic manipulations, the final solution requires the calculation of the matrix inverse, which can be unstable and unobtainable when the number of nodes in a graph is very large.
This first limitation can potentially be mitigated by replacing the spatial smoothness in \eqref{eq_offline_cost_bad} with the Sobolev Smoothness, which captures the smoothness in both the spatial domain and the temporal domain \cite{giraldo_2022_reconstruction_Sobolev_smoothness}.
To address the second limitation, one can approximate the matrix inverse using Newton's method, which gives us an iterative method that can be implemented in a distributed manner \cite{Zhou_2022}.

\subsubsection{Joint Time and Graph Fourier Transform} The foundational work on graph wavelets \cite{HAMMOND2011_wavelet} laid the groundwork for spectral analysis on static graph signals, which later evolved into more comprehensive spatiotemporal frameworks. 
To handle scenarios where graph signals evolve, it is essential to consider both spatial and temporal dimensions simultaneously. 
This has led to the development of Time-vertex methods and the Joint Time and Graph Fourier Transform (JFT) that integrates time-domain information directly into the spectral operations seen in the GSP framework \cite{Loukas_2016_Frequency_JFT}. 
The Joint Time-Vertex Fourier Transform captures both spatial domain variations and temporal domain variations by combining GFT and DFT, providing a comprehensive analysis of spatiotemporal signals and merging classical signal processing and GSP \cite{Grassi_2018_time_vertex,kartal_2022_joint_time_vertex}. 
The JFT also made the definition of dynamic wavelets on graphs possible for tracking the evolving TVG signals \cite{Grassi_2016_Tracking}.
This allows it to effectively decompose signals into joint time-vertex frequency components, offering insights into both temporal dynamics and graph-structured relationships. 
An additional regularization term is introduced to ensure stability and performance,  promoting smoothness across both graph and time dimensions \cite{kartal_2022_joint_time_vertex}. 

\subsubsection{Time-series Analysis on Graphs} As we explained earlier, conventional multivariate time series models, such as Vector Autoregressive (VAR) and Vector Autoregressive Moving Average (VARMA) processes, can be applied directly to graph signals since these signals are inherently multivariate \cite{box2015time}.
However, conventional multivariate time series models are primarily designed for temporal tasks, which limits their ability to exploit the spatial relationships captured within the graph topologies. 
Taking the VARMA model as an example, it has the expression of 
\begin{equation}
    \mathbf{X}[t] = \sum_{j=1}^{J} \boldsymbol{\Phi}_j \mathbf{X}[t-j] + \sum_{k=0}^{K} \boldsymbol{\Theta}_k \boldsymbol{\Xi}[t-k],
    \label{eq_VARMA}
\end{equation}
where $J$ is the AR order, $K$ is the MA order, $\mathbf{X}[t]$ is a multi-variate random variable, $\boldsymbol{\Phi}_i$ is the $N$ by $N$ AR parameter, and $\boldsymbol{\Theta}_j$ is the $N$ by $N$ MA parameter, $\boldsymbol{\Theta}_0$ is an identity matrix, and $\boldsymbol{\Xi}[t]$ denotes the matrix of white noise at time $t$.
It becomes difficult to accurately infer model parameters in \eqref{eq_VARMA} when there are nodes consistently missing, causing conventional VARMA models to underperform.
To address temporal variability and capture space-time interactions, time series analysis is integrated with GSP.

Time-series analysis algorithms were introduced into GSP by specifying their parameters in the spectral domain via GFT, using the multivariate time-series as graph signals and potentially decreasing parameter size compared to traditional VAR and VARMA \cite{Mei_GVAR_2017, Isufi2019gvarma}. 
The GSP analogy of VAR and VARMA is the graph Vector Autoregressive (GVAR) model \cite{Mei_GVAR_2017} and the graph Vector Autoregressive–Moving-Average (GVARMA) model \cite{Isufi_GVARMA_2017}, respectively. 
Specifically, in VAR, the temporal association is modeled by an AR process, and the spatial association is captured by describing the matrix coefficients of the AR process as graph polynomial filters. 
In GVARMA, the VARMA coefficients are designed independently from the underlying graph, but instead are spectral filters. 
The filter design now takes into account both the spatial and temporal domains:
\begin{equation}
    \mathbf{X}[t] = \sum_{j=1}^{J} \boldsymbol{\Phi}_j(\mathbf{L}) \mathbf{X}[t-j] + \sum_{k=0}^{K} \boldsymbol{\Theta}_k(\mathbf{L}) \boldsymbol{\Xi}[t-k].
    \label{eq_GVARMA}
\end{equation}
In \eqref{eq_GVARMA}, to incorporate the spatial information  through GSP, the parameters are now defined using graph convolution \eqref{eq_spectral_graph_convolution}:
\begin{equation}
        \boldsymbol{\Phi}_j(\mathbf{L}) = \mathbf{U}\mathbf{H}_{\Phi}(\boldsymbol{\Lambda})\mathbf{U}^\top \text{ and }
    \boldsymbol{\Theta}_k(\mathbf{L}) = \mathbf{U}\mathbf{H}_{\Theta}(\boldsymbol{\Lambda})\mathbf{U}^\top.
    \label{eq_GVARMA_parameter_spectral}
\end{equation}
Additionally, these parameters can also be defined by spatial graph methods using the Chebyshev approximation \eqref{eq_cheb_H_1} \cite{Isufi2019gvarma}:
\begin{equation}
        \boldsymbol{\Phi}_j(\mathbf{L}) = \sum_{p=0}^{P}{\phi}_{p, j} \mathbf{L}^p \text{ and } \boldsymbol{\Theta}_k(\mathbf{L}) = \sum_{q=0}^{Q}{\theta}_{q, k} \mathbf{L}^q.
        \label{eq_GVARMA_parameter_spatial}
\end{equation}
Parameter reduction in GVARMA can be achieved as follows. 
Even though VARMA and GVARMA parameters have size $N$ by $N$, GVARMA  \eqref{eq_GVARMA_parameter_spectral} defines the parameters in the spectral domain as shown in \eqref{eq_GVARMA_parameter_spectral} by using the filter $\mathbf{H}(\boldsymbol{\Lambda})$, which has only $N$ parameters.
The number of parameters can be further reduced by defining a frequency set $\mathcal{F}$ with $|\mathcal{F}|<N$, and then defining parameters only for the frequencies that are present in $\mathcal{F}$.
As a result, the actual size of the parameters the GVARMA needs in \eqref{eq_GVARMA_parameter_spectral} is drastically reduced to $|\mathcal{F}| << N^2$.
Similarly, if the GVARMA was to be derived from the spatial domain as shown in \eqref{eq_GVARMA_parameter_spatial}, the actual sizes of the parameters that the GVARMA needs to learn are $P << N^2$ and $Q <, N^2$ in $\boldsymbol{\Phi}_j$ and $\boldsymbol{\Theta}_k$ respectively, for all $j = 1\cdots J$ and $K = 1 \cdots K$. 

VARMA and GVARMA models are limited by their assumption of constant covariance, which makes them insufficient for capturing time-varying features such as volatility. 
Volatility, a key concept in finance, has been extensively studied by researchers. In time series analysis, models such as Autoregressive Conditional Heteroscedasticity (ARCH) \cite{engle1982autoregressive} and its generalized version (GARCH) \cite{bollerslev1986generalized} were proposed to address symmetric volatility. 
To capture asymmetric volatility, the exponential GARCH (EGARCH) model \cite{nelson1991conditional} and the GJR-GARCH model \cite{glosten1993relation} were subsequently developed.
Recently, a unified framework for defining GARCH models on graphs was introduced \cite{Hong_GGARCH_2023}. 
The G-GARCH model shares the same model formula as GVARMA in \eqref{eq_GVARMA}, with the exception that the covariance matrix follows a GVARMA model instead of constant variance.
By leveraging the graph spectral domain, G-GARCH decomposes a multivariate GARCH model into a linear combination of several univariate GARCH processes, enabling efficient processing of graph-structured data.
The G-GARCH model can also be derived into asymmetric models such as the G-EGARCH and G-GJR-GARCH \cite{Hong_GGARCH_2023}. 

\textbf{Remarks} Offline algorithms are most preferred when the long temporal sequences in dataset 
is available prior to processing and when long-term temporal dependencies need 
to be exploited. 
By accessing the longer temporal horizon, offline TVGSP methods 
can leverage richer information compared with their online counterparts. 
However, this capability comes with increased 
computational and memory requirements, and offline processing may be unsuitable 
in scenarios that demand immediate decision making or involve rapidly evolving 
graph signals. 
Therefore, offline TVGSP algorithms are best suited for 
applications prioritizing accuracy, global temporal understanding, and 
algorithmic expressiveness over real-time responsiveness.

\section{Neural Networks for Time-Varying Graphs}
\label{section_GNN}
Augmenting GSP models with nonlinear and adaptive components, instead of redesigning them from scratch, offers a principled path to handling richer dependencies and temporal dynamics. 
Early GNNs follow this philosophy by combining graph polynomial filters such as those in \eqref{eq_cheb_H_1} with trainable weights and nonlinear activation functions, achieving strong performance on tasks including node classification, link prediction, and node regression~\cite{bruna2013_spectral_GCN, defferrard2016_cheb, kipf2017semi_supervised, chamberlain_2022_graph}.
However, these GNNs are designed for static graphs.
When applied to time-varying signals, they usually operate on a single static snapshot, largely ignoring the temporal dependencies in the data. 
As a result, their performance can degrade on tasks that rely on temporal structure, such as forecasting or dynamic pattern detection. 
This limitation has motivated a rich body of work on GNN architectures that explicitly model and adapt to TVG signals and topologies.
It is worth noting that this section does not aim to provide an exhaustive survey of dynamic GNNs; instead, we highlight representative models that are most relevant to time-varying GSP and the perspective developed in this work. 
For completeness, we also discuss continual-learning-based DGNNs and CTDG methods. 
We emphasize that corresponding developments in TVGSP remain scarce; our goal here is therefore not to establish a one-to-one correspondence, but to identify important problem regimes where TVGSP may be extended in the future.

\subsection{Spatiotemporal Graph Neural Networks (STGNNs)}
\label{section_STGNN}
\subsubsection{Decoupled Learning Methods}
Learning spatial and temporal features, respectively, is the most natural idea when dealing with TVGs. 
Based on this idea, several decoupled spatiotemporal GNNs are proposed~\cite{yu2018_STGCN, zhao2019tgcn, zheng2020gman, Ruiz_2020_gated}. 
The Spatiotemporal Graph Convolutional Network (STGCN)~\cite{yu2018_STGCN} is a typical decoupled learning method, in which spatial and temporal features are processed successively. 
The layer-wise propagation rule can be represented as
\begin{equation}
    \mathbf{Z}^{(l+1)}
    =
    \Gamma_1^{(l)} *_\mathcal{T}
    \,
    \text{ReLU}\!\Big(
        \Theta^{(l)} *_\mathcal{G}
        \big(
            \Gamma_0^{(l)} *_\mathcal{T} \mathbf{Z}^{(l)}
        \big)
    \Big),
    \label{eq_STG_GNN}
\end{equation}
where $*_\mathcal{T}$ and $*_\mathcal{G}$ denote temporal and graph convolution operators, respectively, $\sigma(\cdot)$ is the non-linear activation function, and $\mathbf{Z}^{(l)}$ is the spatiotemporal feature tensor at layer $l$. 
The input to the first layer is typically a temporal window of graph signals,
$\mathbf{Z}^{(0)} = \mathbf{X} = \big[\boldsymbol{x}[T_1],\dots,\boldsymbol{x}[T_2]\big]$, 
where $\boldsymbol{x}[t]\in\mathbb{R}^N$ collects the node signals at time $t$ for an $N$-node graph, and $[T_1,T_2]$ denotes a historical time interval used for training. 

Each temporal operator $\Gamma$ is implemented as a gated temporal convolution. 
Specifically, $\Gamma *_\mathcal{T}\mathbf{Z}$ is rooted from two parallel 1D (non-graph) temporal convolutions that produce a feature map $\mathbf{Z}_1$ and a gating map $\mathbf{Z}_2$.
The final temporal output is then obtained via element-wise modulation $\mathbf{Z}_1 \odot \sigma(\mathbf{Z}_2)$. 
For the graph component, STGCN adopts the spectral graph convolution framework, where the learnable filter is parameterized as a polynomial of the normalized Laplacian. In particular, $\Theta *_\mathcal{G} \mathbf{Z} = \Theta(\tilde{\mathbf{L}})\mathbf{Z}$ is implemented using a Chebyshev polynomial approximation exactly as shown in \eqref{eq_cheb_H_1}, yielding an efficient and localized $K$-hop filtering operation. 
Combining these instantiations gives
\begin{equation}
\begin{aligned}
    \Gamma *_\mathcal{T} \mathbf{Z}
        &= \mathbf{Z}_1 \odot \sigma(\mathbf{Z}_2), \\
    \Theta *_\mathcal{G} \mathbf{Z}
        &= \Theta(\tilde{\mathbf{L}})\mathbf{Z}
        \;\approx\;
        \sum_{k=0}^{K-1} \theta_k T_k(\tilde{\mathbf{L}})\mathbf{Z},
\end{aligned}
\label{eq_decoupled}
\end{equation}
where $\mathbf{Z}_1$ and $\mathbf{Z}_2$ denote the two temporal feature maps, $\tilde{\mathbf{L}}$ is the normalized Laplacian, $\theta_k$ are learnable coefficients, and $T_k(\cdot)$ denotes the $k$-th order Chebyshev polynomial~\cite{defferrard2016_cheb}. In STGCN, the temporal operators $\Gamma_0^{(l)}$ and $\Gamma_1^{(l)}$ in~\eqref{eq_STG_GNN} take this gated-convolution form commonly found in non-graph NNs, while $\Theta^{(l)}$ corresponds to the Chebyshev graph filter. 
Unlike the adaptive TVGSP filters discussed in Section~\ref{sec_GSP}, which can operate online as new observations arrive, STGCN and related STGNNs are typically trained offline on historical windows $\{\boldsymbol{x}[T_1],\dots,\boldsymbol{x}[T_2]\}$ and then used for prediction under an implicit (quasi-)stationarity assumption, highlighting a key conceptual difference between learning-based STGNNs and adaptive GSP-based approaches.

From a GSP viewpoint, the spatial convolution in~\eqref{eq_decoupled} is equivalent to a graph spectral filtering operator,
\begin{equation}
    \Theta *_\mathcal{G} \mathbf{Z}
    = \mathbf{U}\,\mathbf{H}(\boldsymbol{\Lambda})\,\mathbf{U}^\top \mathbf{Z}
    \;\approx\;
    \Big(\theta_0\mathbf{I} + \sum_{k=1}^{K-1}\theta_k\,T_k(\tilde{\mathbf{L}})\Big)\mathbf{Z},
    \label{eq_stgcn_filter_gsp}
\end{equation}
which directly parallels the Chebyshev polynomial approximation in~\eqref{eq_cheb_H_1}.
This interpretation establishes a direct theoretical correspondence between STGNN and GSP, showing that the spatial component of STGCN can be explicitly interpreted as a learnable GSP filter in the spectral domain, thereby providing interpretability while retaining neural flexibility.    

\subsubsection{Integrated Learning Methods} Compared with the successive learning of spatial and temporal features by two independent modules, merging spatial and temporal feature learning into one layer can be more conducive to learning joint information. Motivated by this, a series of integrated spatiotemporal GNNs are proposed~\cite{li2017diffusion, wang2021modeling, huang2021spatio}, where a GCN interpretable by GSP in \eqref{eq_cheb_H_1} is employed as a spatial feature extractor in an RNN layer for joint spatiotemporal learning. 
For example, the Diffusion Convolutional Recurrent Neural Networks (DC-RNN) was proposed to insert a tailored diffusion convolution into each RNN layer~\cite{li2017diffusion}:
\begin{equation}
    \boldsymbol{\Theta} *_\mathcal{G} \tilde{\boldsymbol{z}} = \sum_{k=0}^{K-1}(\theta_{k,1}(\mathbf{D}_O^{-1}\mathbf{W})^k + \theta_{k,2}(\mathbf{D}_I^{-1}\mathbf{W}^\top)^k)\boldsymbol{z},
\end{equation}
where $\theta_{k,1}$ and $\theta_{k,2}$ are two independent parameters for the transition matrices of diffusion process $\mathbf{D}_O^{-1}\mathbf{W}$ and reverse one $\mathbf{D}_I^{-1}\mathbf{W}^\top$ respectively. 
Then, the diffusion convolution with activation function is used to replace the matrix multiplications in the Gated Recurrent Units (GRU) to construct a Diffusion Convolution GRU (DCGRU) as a basic unit in its sequential model, processing temporal information and spatial information collaboratively.
From a GSP viewpoint, this corresponds to 2 separate graph filtering operations representing the forward and reverse diffusion processes
\begin{equation}
\begin{aligned}
    \boldsymbol{\Theta} *_\mathcal{G} \tilde{\boldsymbol{z}}
    &= \underbrace{\sum_{k=0}^{K-1}\theta_{k,1}(\mathbf{D}_O^{-1}\mathbf{W})^k\tilde{\boldsymbol{z}}}_{\text{forward diffusion}}
     \;+\;
       \underbrace{\sum_{k=0}^{K-1}\theta_{k,2}(\mathbf{D}_I^{-1}\mathbf{W}^\top)^k\tilde{\boldsymbol{z}}}_{\text{reverse diffusion}} \\[2mm]
    &\approx
    \mathbf{U}\,\mathbf{H}_1(\boldsymbol{\Lambda})\,\mathbf{U}^\top \tilde{\boldsymbol{z}}
    \;+\;
    \mathbf{U}\,\mathbf{H}_2(\boldsymbol{\Lambda})\,\mathbf{U}^\top \tilde{\boldsymbol{z}},
\end{aligned}
\label{eq_dc_forward_backward}
\end{equation}
which makes the spatial processing in DC-RNN explicitly interpretable as a learnable spectral filtering process, while the recurrent structure simultaneously models temporal evolution. 
This unified interpretation highlights how integrated STGNNs inherently perform joint GSP filtering and temporal state modeling.

\subsubsection{Noise-Aware Methods} Some methods also take into account the noise-induced data missing in modeling, and the representative work is the spatiotemporal Graph Gated Recurrent Units (SG-GRU)~\cite{Lewnfus_2020_Joint} that are developed to make predictions temporally and complete the missing nodes simultaneously with the aid of GFT and GRU. To achieve this, it first calculates the graph interpolation operator $\Phi_S$ with a given admissible graph sampling operator $\Psi_S$:
\begin{equation}
    ||\boldsymbol{x}-\Phi_S\Psi_S\boldsymbol{x}||_2 \leq \frac{\epsilon}{SV_{min}(\Psi_S\mathbf{U}_{:,\mathcal{F}})},
\end{equation}
where $SV_{min}(\cdot)$ means the minimal singular value of the given matrix; $\mathbf{U}_{:,\mathcal{F}}$ is the sub-matrix of the eigenvector matrix $\mathbf{U}$ with columns restricted to the indices associated with the frequencies indicated in $\mathcal{F}$; $\boldsymbol{x}$ is assumed as $(\mathcal{F},\epsilon)$ bandlimited. 
By this constraint, the graph interpolation operator $\Phi_S$ is used to recover the corrupted graph signal after forecasting:
\begin{equation}
    \tilde{\boldsymbol{x}}^{(t+1)} = \text{MLP}(\Phi_S\hat{\boldsymbol{y}}^{(t+1)}, \Phi_S\mathbf{U}_{:,\mathcal{F}} \hat{\boldsymbol{z}}^{(t+1)}),
\end{equation}
where MLP stands for multi-layer perceptron, $\hat{\boldsymbol{y}}^{(t+1)}$ and $\hat{\boldsymbol{z}}^{(t+1)}$ are predicted by GRU and Spectral GRU from $\boldsymbol{x}^{(t)}$ respectively. The major drawback of this empirical approach is that the GFT only acts as a domain transformation, and the performance relies only on the GRU, which is not adapted to the graphs and lacks interpretability. Besides, the hyperparameters selection (\eg $\Psi_S$, $\epsilon$, and $\mathcal{F}$) heavily depends on the prior knowledge of the graph.

\textbf{Remarks:} 
Constructing an STGNN to collaboratively learn and process the time-varying features is a direct way to cope with time-varying challenges.
The STGNNs mostly resemble the TVGSP approaches, such as the spatial graph convolution seen in \eqref{eq_spatial_GCN} or \eqref{eq_spectral_GCN}, with the inclusion of learnable parameters and activation functions. 
As shown in (a) of Fig.~\ref{fig:tvgnn}, the paradigm of STGNNs can be summarized as consisting of a pair of time modules (\eg 1D convolution, RNN and GRU) and spatial modules (\eg GCN, ChebNet, and GAT), and an extra spatiotemporal fusion module. 
From a GSP perspective, the spatial modules in these GNNs can be interpreted as graph filtering operations implemented via the GFT, where learned parameters implicitly define spectral responses over the graph.
The difference between various STGNNs generally appears in the architecture design of these modules and how spatial and temporal features are fused. 
Differences across STGNN variants primarily arise from how these modules are architected and how spatial and temporal information is fused, while some approaches additionally incorporate denoising or inpainting mechanisms during spatiotemporal learning~\cite{castro2023time}. This close relationship suggests promising opportunities for transferring GSP insights on temporal regularization, filter design, and robustness into the design and interpretation of dynamic GNNs.

\subsection{Dynamic Graph Neural Networks}
\label{sec_DGNN}
Most of the STGNNs are designed for node-level time-varying signal processing, while the temporal variation on higher-order signals (\eg edges and topologies) is overlooked. Recently, Dynamic Graph Neural Networks (DGNN) were proposed to address this problem. In DGNNs, the TVG is divided into DTDGs and CTDGs, the two classes of dynamic graphs that correspond to the categories we defined in Section~\ref{sec_background}. 
By different dynamic graph modeling, various DGNNs are introduced. We then discuss them separately.

\subsubsection{DTDG Methods} As illustrated in Fig.~\ref{fig:tvgnn}(b), typical DGNNs for DTDG processing follow an RNN-inspired pipeline. At each discrete timestamp $t$, a spatial GNN extracts node representations from the current graph snapshot $\mathcal{G}[t] = (\mathbf{L}[t], \mathbf{X}[t])$, and a temporal module propagates information over time, i.e.,
\begin{equation}
    \mathbf{Z}[t] = g(\mathcal{G}[t]), \qquad
    \mathbf{S}[t] = f\big(\mathbf{S}[t-1], \mathbf{Z}[t]\big),
\end{equation}
where $\mathbf{X}[t]$ denotes the input graph signal (node features), $\mathbf{Z}[t]$ denotes the learned node representations at time $t$, and $\mathbf{S}[t]$ is a hidden state capturing temporal dependencies. 
In practice, $g(\cdot)$ is typically a GCN-inspired structure implemented using $\mathbf{L}[t]$, while $f(\cdot)$ is often implemented as a recurrent or linear projection module. Unlike STGNNs with fixed topology, DTDG models explicitly leverage the evolving graph structure at each timestamp to better capture time-varying node relationships~\cite{manessi2020dynamic, li2019predicting}. 
EvolveGCN is another DTDG-GNN framework in which an RNN dynamically updates the GCN parameters across discrete graph snapshots, enabling effective learning on evolving DTDGs~\cite{pareja2020evolvegcn}.
Variational Graph Recurrent Neural Networks model TVGs as DTDGs using a variational autoencoder framework, in which latent node representations evolve through a recurrent prior across successive graph snapshots, enabling effective representation learning and link prediction ~\cite{hajiramezanali2019variational}.
These models are representative of GNNS using DTDGs; here, we highlight them as key examples, while recognizing that the broader DGNN literature is considerably richer.

From a GSP perspective, the function $g(\mathcal{G}[t], \mathbf{X}[t])$ in the above DTDG pipeline can be interpreted as applying a TVG filter to the node features at each time step. 
In particular, our GSP formulation in Section~\ref{sec_GSP} considers time-varying spatial operators of the form~\eqref{eq_adaptive_spatial_weight}. 
When such an operator is used as the spatial component of $g(\cdot)$, a natural GSP-inspired realization is
\begin{equation}
    \mathbf{Z}[t]
    = g(\mathcal{G}[t], \mathbf{X}[t]) 
    = \sigma\big(\mathbf{W}[t]\mathbf{X}[t]\big),
    \label{eq_dtdg_gsp_bridge}
\end{equation}
where $\mathbf{X}[t]$ denotes the node features at time $t$ and $\sigma(\cdot)$ is a pointwise nonlinearity. In this view, many DTDG GNNs implicitly learn a sequence of TVG filters $\mathbf{W}[t]$ through their parameters, whereas the GSP-based formulation in~\eqref{eq_adaptive_spatial_weight} makes these filters explicit in terms of powers of the time-varying Laplacian $\mathbf{L}[t]$ that are directly determined by the evolving topology $\mathcal{G}[t]$. 
This correspondence provides a principled bridge between DTDG GNN architectures and TVGSP, and suggests opportunities to transfer stability and regularization insights from GSP to the design of dynamic GNNs.

\subsubsection{CTDG Methods} Since the graph representation is different from DTDG, where the event set stores the temporal information rather than iteratively forwarding previous information like RNN. As a consequence, the DGNN methods for CTDG are more diverse. A class of methods models the event sequence by a temporal point process, where the occurrence of events recording the evolving graph signal is assumed to follow a multivariate temporal point process. The conditional intensity function of this process can be modulated by the score of the relevant relationship, and the measurement of the relationship can be realized by GNNs~\cite{trivedi2017know, trivedi2019dyrep}. Besides the temporal point process, as an extension of random walk~\cite{perozzi2014deepwalk,grover2016node2vec}, the temporal random walk is utilized for learning spatiotemporal association as well. In a temporal random walk, the movement from one vertex to another will depend on its spatiotemporal neighbors to extract dynamics from the TVG~\cite{wang2021inductive,li2023zebra}.

On the other hand, some researchers also model the relationship between temporal events directly by neural networks. For instance, the Dynamic Graph Neural Networks (DyGNN)~\cite{ma2020DGCN} employ an update component and a propagation component to cope with coming events. As shown in (c) of Fig.~\ref{fig:tvgnn}, when a new event comes, the update component maintains the freshness of node information by learning the relationship with previously memorized events and the time interval between the stored events and the new event. Then, the propagation component propagates the updated information to the affected nodes (\eg its neighborhood nodes) by considering the intensity of the influence. 
Meanwhile,  JODIE~\cite{kumar2019JODIE} is developed and shares a similar scheme with DyGNN, which also has the update operation and the projection operation. The major difference is that in JODIE, each node has a static embedding to represent its fixed attributes and a dynamic embedding to reflect its current state.


\subsection{Graph Lifelong Learning}
Graph lifelong learning~\cite{wang2022lifelong} (a.k.a., continual graph learning~\cite{zhou2021overcoming}) is an emerging research topic and is mainly developed for processing real-world time-growing graphs. The scene of graph lifelong learning is that the number of nodes is growing over time, or there will be new nodes of unseen classes. As depicted in (d) of Fig.~\ref{fig:tvgnn}, it solves this problem by iteratively training the GNN. Like general continual learning, the core problem in graph lifelong learning is also catastrophic forgetting. To prevent this, the current graph lifelong learning methods mostly refers to continual learning and can be classified into three categories, including the architectural methods, the regularization methods, and the experience-replay methods. 

\subsubsection{Architectural Methods} The architectural methods~\cite{wang2022lifelong,sun2019hierarchical} aim to change the architecture of the GNN iteratively to coherently adapt to the dynamic data stream. The Feature Graph Network~\cite{wang2022lifelong} is a special type of architectural method to tackle time-growing graphs by node feature transformation. Specifically, it transforms each node into a feature graph, and the features in this node become new nodes in each feature graph. The edge weights between these feature nodes are determined by cross-correlation. Specifically, if the feature node matrix is $\mathbf{X}\in \mathbb{R}^{F\times C}$, the feature adjacency matrix is
 \begin{equation}
\mathbf{A}_{k,c}^f(\mathbf{X})=\text{sgnroot}\left(\frac{\sum_{\mathbf{Y}\in N_{k}(\mathbf{X})} \boldsymbol{w}_{X,Y}\mathbf{X}[:,c]\mathbf{Y}[:,c]^\top}{|N_k(\mathbf{X})|}\right),
 \end{equation}
where $k$ represents the $k$-hop neighbor of $\mathbf{X}$, $c$ indexes the channel, and $N_k(\mathbf{X})$ is the number of $k$-hop neighbor of $\mathbf{X}$. In this way, the continually coming nodes are regarded as an individual graph and classified with previous techniques in continual learning.
 
\subsubsection{Experience-Replay Methods} In addition to architectural methods, like in continual learning, the experience replay technique is also employed for graph continual learning~\cite{liu2021overcoming,galke2021lifelong}. In the Experience Replay Graph Neural Networks (ER-GNN)~\cite{liu2021overcoming}, a set of nodes is stored in a buffer as experience and replayed in the subsequent task to avoid forgetting. To identify valuable nodes of each task, ER-GNN proposes several metrics, including the mean of features, coverage maximization, and influence maximization. Once the nodes are selected and stored in a buffer, the training objective for each task is altered as:
\begin{equation}
\begin{split}
    \mathcal{L}_{k}'(f_\theta, D_k, B) &= \beta\mathcal{L}_{k}(f_\theta, D_k) + (1-\beta)\mathcal{L}_{k}(f_\theta, B), \\
    \beta &= |B|/(|D_k|+|B|),
\end{split}
\end{equation}
where $\mathcal{L}$ indicates the loss function (\eg cross entropy in node classification tasks); $k$ indexes the task; $B$ represents the experience buffer; $f_\theta$ denotes the GNN; $D_k$ represents the training dataset of task $k$.

\textbf{Regularization Methods} Adding regularization terms to penalize dramatic updates on important parameters is common in continual learning~\cite{zhou2021overcoming, tan2022graph,kou2020disentangle}. The representative one in this line of work is a plug-and-play module called Topology-aware Weight Preserving~\cite{zhou2021overcoming}, which estimates an importance score for each network parameter. This score is computed as the sum of gradients concerning parameters (denoted as $I^{loss}$) and gradients concerning attention coefficients parameters (denoted as $I^{(ts)}$). The first term aims to minimize the loss of different tasks, and it can be calculated from:
\begin{equation}
\begin{split}
    \mathcal{L}(\mathbf{X}_k, \mathbf{W}+\Delta_\mathbf{W}) - \mathcal{L}(\mathbf{X}_k, \mathbf{W}) &= \sum_m f(\mathbf{X}_k)\Delta \boldsymbol{w}_m, \\
    I_k^{loss} &= ||f(\mathbf{X}_k)||_2,
\end{split}
\end{equation}
where $\mathbf{X}_k$ is the training feature embedding of task $k$; $\mathbf{W}=\{\boldsymbol{w}_m\}$ is the network parameter; $\mathcal{L}$ is the loss function. For the second term, it is calculated from:
\begin{small}
    \begin{equation}
\begin{split}
    a(\mathbf{F}_{i,j}^{(l-1)};\mathbf{W}^l + \Delta_\mathbf{W}^l) - a(\mathbf{F}_{i,j}^{(l-1)};\mathbf{W}^l) &= \sum_m g(\mathbf{F}_{i,j}^{(l-1)})\Delta \boldsymbol{w}_{m},\\
    I_k^{(ts)} &= ||g(\mathbf{F}_{i,j}^{(l-1)})||_2,
\end{split}
\end{equation}
\end{small}
where $a$ represents a single neural network layer for feature projection, $f(\cdot)$ and
$g(\cdot)$ are gradient losses, $\mathbf{F}_{i,j}^{(l-1)}$ represents the concatenated feature embedding of node $i$ and $j$ at $(l-1)^\text{th}$ layer in the graph attention neural network.
Dealing with the TVG by graph continual learning is a noteworthy direction, which merges techniques from both continual learning and GNNs and achieves great performance in some specific scenarios (\eg learning and processing the time-growing graphs~\cite{wang2022lifelong}).

\section{Applications}
\label{sec_application}

On the application side, recent advancements in TVGSP and TVGNNs have driven progress in various fields, including social science, biological analysis, transportation, environmental studies, and finance. This growth is attributed to the exceptional capability of TVG representation in capturing dynamic relationships among irregularly distributed signals \cite{Sandryhaila_2014_big_data, Shuman_2013_the_emerging, Ortega_graph_2018, Dong_Graph_ML_2020}.
Here, we will be providing a discussion on how TVGSP and TVGNN algorithms are utilized.

\subsection{Social Science}
In social science, TVGSP and TVGNN emerged as powerful tools for analyzing dynamic datasets, offering insights into evolving relationships, behaviors, and social interaction patterns. Applications in this domain span various areas, such as opinion dynamics, community detection, anomaly detection, and modeling social behavior.

\subsubsection{Opinion Dynamics and Influence Propagation:} Opinion dynamics and influence propagation in social networks can be modeled using TVGs, where nodes represent users and edges represent temporal interactions. TVGs enable tracking the spread and evolution of opinions or sentiments. Dynamic graph embeddings on platforms like Reddit and Wikipedia showcase this \cite{shetty2004enron, kumar2019predicting}.

\subsubsection{Community Detection:}  
TVGSP enables the monitoring of changes, mergers, or splits of communities in social networks. 
Community detection in dynamic networks benefits greatly from TVG-based methodologies, which allow for the identification and tracking of evolving communities over time in physical and digital contexts.
For example, TVG analysis enables researchers to detect emerging communities and shifting user preferences over time, providing valuable insight into online social interactions~\cite{panzarasa2009patterns, sapiezynski2019interaction}.

\subsubsection{Detection of Anomalous Patterns:} TVGSP is highly useful for anomaly detection in social networks, where unusual patterns in user behavior or interactions can be flagged in real time. Such capabilities are crucial for monitoring social media for events like coordinated misinformation efforts or network breaches. 
In the financial sectors, the DGraphFin \cite{huang2022dgraph} dataset supports such TVGSP applications, where anomaly detection algorithms are used to detect fraud and irregular transaction patterns. In social contexts, \cite{beres2018temporal} introduces a centrality measure designed for continuously updating network streams, helping identify influential nodes and interactions as new data arrives. \cite{bridges2016multi} employed a multi-level anomaly detection algorithm in dynamic graph settings, effectively detecting anomalies at different graph levels (\eg node, subgraph) with interactive visualization. These methods prove valuable for anomaly detection within evolving social graphs where central actors can be continually monitored.

\subsection{Biomedical analysis} Dynamic graph-based methods are particularly suited to represent temporally evolving interactions, such as neural connectivity patterns, gene expression, and physiological signals. Biomedical applications of TVGSP encompass brain connectivity studies, predictive modeling in medical diagnostics, and protein interaction analysis.

\subsubsection{Dynamic Functional Brain Connectivity Analysis:} Functional connectivity analysis is an increasingly important method for studying brain function, examining the statistical codependencies between signals from different brain regions~\cite{hou2025brainnetmlp}. TVG has become a valuable representation for analyzing functional brain connectivity, particularly in studies using time-series data from fMRI or EEG. 
Yu et al. \cite{yu2015assessing} used TVGs to assess dynamic brain connectivity in patients with schizophrenia by analyzing time-varying functional networks derived from fMRI data, highlighting how TVG-based approaches identify connectivity alterations associated with mental disorders. 
Calhoun and Adali further expanded by using TVG-based techniques to track connectivity states across different brain regions, demonstrating applications in both healthy and clinical populations \cite{calhoun2016time}. 
Zhao et al. achieved superior performance in brain disorder classification by integrating a particle filtering algorithm with GNNs \cite{zhao2024sequential}. Hou et al. integrated the generation, clustering and classification into one framework by learning the spectral dynamics and optimal clustering \cite{hou2025losc}

\subsubsection{Medical Diagnostics Modeling:} In medical diagnostics, TVGSP and TVGNNs have enabled predictive modeling using dynamic patient data, offering tools for early diagnosis and outcome prediction. Xie et al. \cite{xu2024predicting} introduced a graph convolutional neural network that predicts intensive care interventions based on temporal multivariate data from ICU records, demonstrating the potential for real-time decision support in critical care environments. 
Li et al. \cite{hasan2021application} explored TVGs in EEG analysis to monitor cognitive health and diagnose neurological disorders, showing that dynamic graphs can reveal early-stage signs of brain dysfunction in clinical populations.

\subsubsection{Protein Interaction and Structural Dynamics Analysis:} TVGs provided a robust approach for studying dynamic protein interactions, enabling the modeling of temporal and structural changes within protein-protein interaction (PPI) networks. By applying TVGSP and TVGNNs, researchers can capture the evolution of interactions among proteins, which is essential for understanding functional pathways and responses to various stimuli. 
Song et al. \cite{song2009time} applied dynamic Bayesian networks to track protein signaling pathways over time, using TVGs to model interactions during cellular processes. 
Li et al. \cite{li2022graph} demonstrated how TVGs can track dynamic changes in PPI networks, highlighting structural adaptations in proteins under different environmental conditions. 
Calazans et al. \cite{calazans2024machine} leveraged TVGSP and TVGNNs for protein interaction analysis, using graph learning techniques to identify structural motifs and interaction dynamics in protein complexes. 

\begin{figure}[htb]
    \centering
    \includegraphics[trim={100 300 50 250},clip,width=\linewidth]{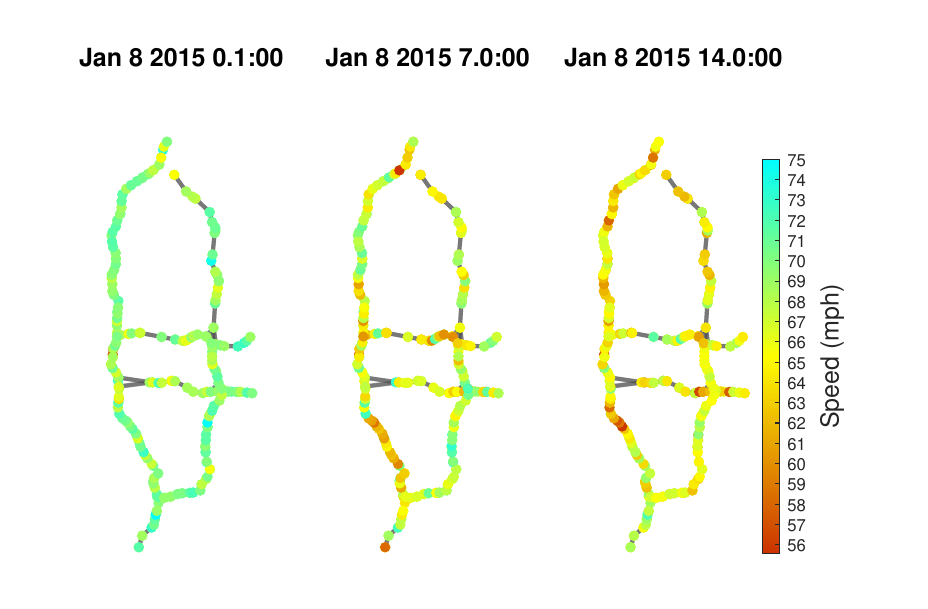}
    \caption{The Seattle Loop dataset at 3 different time instances.}
    \label{fig:seattle}
\end{figure}

\subsection{Transportation}
In transportation systems, TVGs have become an essential tool for analyzing and capturing the complex and evolving relationships between various components, such as vehicles, traffic signals, and road networks. 
Sensors along the roads are often modeled as graph nodes or edges, with the captured sensor readings treated as time-varying signals.
Key applications include traffic flow prediction, congestion management, and route optimization. 

\subsubsection{Traffic Flow Prediction and Congestion Analysis:} Traffic flow prediction is one of the most prominent applications of TVGSP and TVGNNs. 
By treating traffic flow as a signal on a graph, TVG-based models capture both spatial dependencies (interconnectedness of adjacent road segments) and temporal variations (changes in flow over time). 
The widely used PeMS datasets from California, such as PeMSD7 \cite{yu2018_STGCN} and PeMSBay \cite{li2017diffusion}, contain rich traffic sensor data, enabling effective applications of GSP for short- and long-term traffic prediction. 
For example, Li et al. \cite{li2017diffusion} utilized GCNs on the PeMS datasets to accurately forecast traffic flows by learning spatial-temporal dependencies across the network.

\subsubsection{Transportation Network Analysis:} TVGSP and TVGNN methods support the analysis and optimization of public transportation networks, where interactions between transit nodes (\eg bus stops, train stations) and connections (routes) change continuously over time \cite{rozemberczki2021pytorch}. 
By analyzing passenger demand patterns, TVG-based methods can optimize transit scheduling and enhance network efficiency. 
For instance, GPS-based movement data from taxis are analyzed by TVGCNs to extract passenger demand patterns and recommend efficient resource allocation strategies \cite{zhao2019tgcn}. Besides, TVGSP algorithms have also been applied for the analysis of GPS displacement~\cite{piriyasatit2024time}

\subsubsection{Route Optimization and Navigation:} For real-time route optimization, TVG approaches with graphs that have edge weights that vary with traffic conditions, road closures, or accidents, enabling route recommendations and adapting to evolving traffic patterns. 
For instance, STGs are used to model real-time traffic speed data from Los Angeles,  enabling adaptive routing recommendations based on evolving traffic patterns and optimizing travel routes in real time \cite{li2017diffusion}. 
TVGs provide predictive frameworks for identifying probable path failure, supporting route planning that can proactively avoid high-risk paths \cite{li2019predicting}.

\subsection{Environmental Science} Evolving interactions of the sensors on sensor grids across different locations naturally form a graph based on sensor locations.
TVGs serve as valuable frameworks for monitoring and modeling natural processes, with applications in air quality forecasting, Meteorological prediction, and environmental hazard detection.

\subsubsection{Air Quality Prediction:} TVG approaches enable the analysis of air pollution signals, providing real-time insights into pollutant levels. 
A model for reconstructing TVG signals was introduced by Qiu et al. \cite{qiu2017time}, which was applied to PM2.5 pollution data from California. This model enables enhanced tracking of pollution levels over time by leveraging the spatial dependencies between monitoring stations. 
A Sobolev smoothness-based framework demonstrated effective signal recovery in environmental datasets, highlighting its application to air quality data \cite{giraldo2022reconstruction}. 
Spatiotemporal smoothness facilitates predictions of air quality by modeling correlations across sensor networks and adapting to temporal changes \cite{liu2019graph}.

\subsubsection{Weather Forecasting:} TVGs play a crucial role in analyzing Meteorological data.
Castro-Correa et al. \cite{castro2024gegenbauer} advanced this area by developing Gegenbauer GNNs specifically designed for reconstructing time-varying signals in complex climate data, allowing for the extraction of meaningful trends over extended periods. 
There are various meteorological prediction scenarios, such as wind prediction \cite{yan2023fast, Hong_GGARCH_2023, stanczyk2021deep, rathore2021multi} and temperature prediction \cite{Spelta_2020_NLMS,jia2021physics,chen2022physics, yan_2022_sign}, illustrating the capability of TVGSP in practice.

\subsubsection{Environmental Hazard Detection:} Monitoring environmental hazards, such as floods, wildfires, and industrial pollution, benefits from the dynamic adaptability of TVGSP frameworks, which allow for the real-time detection of anomalies and the tracking of evolving hazard patterns. 
Valdivia et al. \cite{valdivia2015wavelet} utilized wavelet-based methods to visualize and process time-varying environmental data, enabling real-time analysis of environmental events such as wildfires. In a more recent study, Sun et al. \cite{sun2024tvgcn} developed the Time-Varying Graph Convolutional Network (TVGCN) model to predict multivariate spatiotemporal series, which is particularly useful in forecasting hazard-related data such as rainfall and water levels in flood-prone areas.

\subsection{Financial Analysis:} The integration of TVGs within financial analysis addresses diverse and complex challenges within dynamic financial systems, where interactions between entities such as stocks, transactions, and institutions change over time. TVGs empower researchers and analysts to harness the temporal and structural dynamics of financial interactions, enabling real-time insights into evolving networks. Financial applications of TVGSP and TVGNNs include fraud detection, market volatility analysis, and stock price forecasting. 

\subsubsection{Fraud and Anomaly Detection:} 
Financial networks are extensively used for identifying irregular transactions and preventing fraud \cite{yuan2025dynamic}. 
Kurman et al. provided a framework for detecting fraudulent users through temporal analysis of user ratings, focusing on anomalies in user behaviors and their interactions within the network \cite{kumar2018rev2}. 
Weber et al. used GNNs on a temporal dataset of Bitcoin transactions to detect and prevent money laundering activities in digital finance, leveraging the Bitcoin and Elliptic dataset to identify laundering activities \cite{weber2019anti}.

\subsubsection{Market Volatility Analysis and Risk Assessment:} 
\begin{figure}[htbp]
\centering
\includegraphics[width=0.9\linewidth]{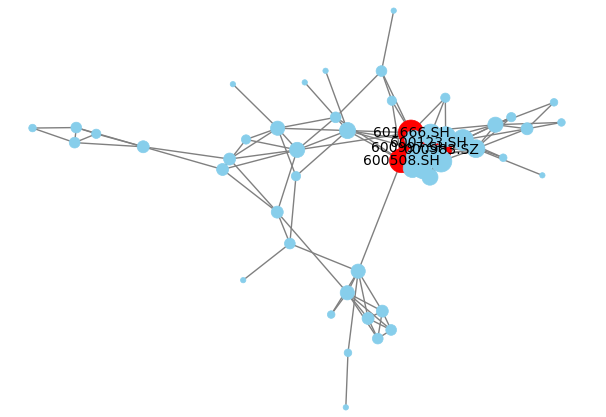}
\caption{Illustrative stock graph of CSI 300 stocks.}
\label{fig:Stock_Graph}
\end{figure}

Predicting stock market volatility often requires monitoring dynamic, interdependent relationships across different stocks or sectors. 
The temporal dynamics captured through TVGs provide insights into correlations and dependencies that influence market risk and volatility. 
For instance, a TVG model can be employed to analyze the dynamic correlations in global stock indices, enabling enhanced forecasting capabilities for market shifts and volatility \cite{chi2024graph}. 
These edge weight prediction techniques help to understand market volatility by focusing on how relationship strengths evolve, offering predictive insights into potential price fluctuations.
The incorporation of topological data analysis tools further elaborates on how changes in the topology and geometry of financial market correlations, particularly during a crash, can be crucial for anticipating drastic market changes and predicting black swan events \cite{yen2021understanding}.
Figure \ref{fig:Stock_Graph} illustrates the correlation among selected CSI 300 stocks, which exemplifies the dynamic correlations that can be analyzed through TVGs.

\subsubsection{Stock Price Prediction:} 
Stock price prediction is a crucial component of financial trading, offering traders valuable insights for making informed decisions on buying, selling, or holding stocks. Accurate forecasts are critical in optimizing trading strategies and maximizing profitability.
A common approach to building the TVG models for stock price prediction is to combine the GNNs (\eg GCN, GAT) with time series predicting methods (\eg RNN, LSTM) \cite{liu2023stock, zheng2023relational}. 

\section{Datasets}
With growing interest in dynamic graphs, various time-varying datasets have emerged for evaluating model performance.
We classify them into three types, STGs, DTDGs, and CTDGs. Here we briefly review the datasets; a summary is in Table~\ref{tab:datasets}.

\subsection{Datasets for STG:} As can be seen from Table~\ref{tab:datasets}, most of the STG datasets belong to traffic data, with a few datasets for epidemic and neuroimaging. 
\textbf{METR-LA} \cite{li2017diffusion} contains data collected from sensors installed on highways in Los Angeles County, capturing traffic speed at various locations at 5-minute intervals. 
\textbf{Seattle Loop} \cite{cui2018deep, cui2019traffic} provides 5-minute interval spatiotemporal speed data collected from inductive loop detectors on Seattle freeways. It features averaged speed measurements from multiple detectors on main lanes in the same direction at specific mileposts. See Fig.~\ref{fig:seattle} for an illustration.
\textbf{Chickenpox} \cite{rozemberczki2021chickenpox} is a spatiotemporal dataset of weekly chickenpox (childhood disease) cases from Hungary. The dataset consists of a county-level adjacency matrix and a time series of the county-level reported cases between 2005 and 2015.
\textbf{HCP} \cite{van2013wu} (Human Connectome Project) is a comprehensive collection of neuroimaging data designed to map the structural and functional connectivity of the human brain from over 1,200 healthy adult participants. 

\subsection{Datasets for CTDG:} The datasets in CTDG cover a wide range of domain types, such as proximity, finance, social, and so on. We chose some representative datasets for each domain to introduce them. 
\textbf{Social Evolution} \cite{li2019predicting} records dynamic interactions among students at MIT's Human Dynamics Lab. Collected via mobile devices, it captures temporal social interactions such as phone calls, text messages, and physical proximity over several months. 
\textbf{Bitcoin-Alpha} \cite{kumar2016edge, kumar2018rev2} is a dynamic network of trust relationships between users on the Bitcoin Alpha platform, where users rate one another based on the trustworthiness of cryptocurrency transactions.
\textbf{LastFM} \cite{kumar2019predicting} contains one month of who-listens-to-which song information, which was timestamped interactions between users and artists on the LastFM music streaming platform, capturing user preferences over time. 
\textbf{UCI} \cite{panzarasa2009patterns} is an online temporal communication, such as emails, between students from the University of California, Irvine, with edges in the graph corresponding to messages exchanged between users.  
\textbf{Taobao} \cite{jin2022neural, zhu2018learning} collected from the Chinese e-commerce platform Taobao, contains temporal interactions between users and items, representing user behavior in an online marketplace, which links the meaning of the type of action (\eg clicking, purchasing, or adding to a wishlist).

\subsection{Datasets for DTDG:} 
\begin{figure*}[ht] 
    \centering
    \begin{subfigure}{0.3\textwidth} 
        \includegraphics[trim={100 300 50 300},clip,width=\linewidth]{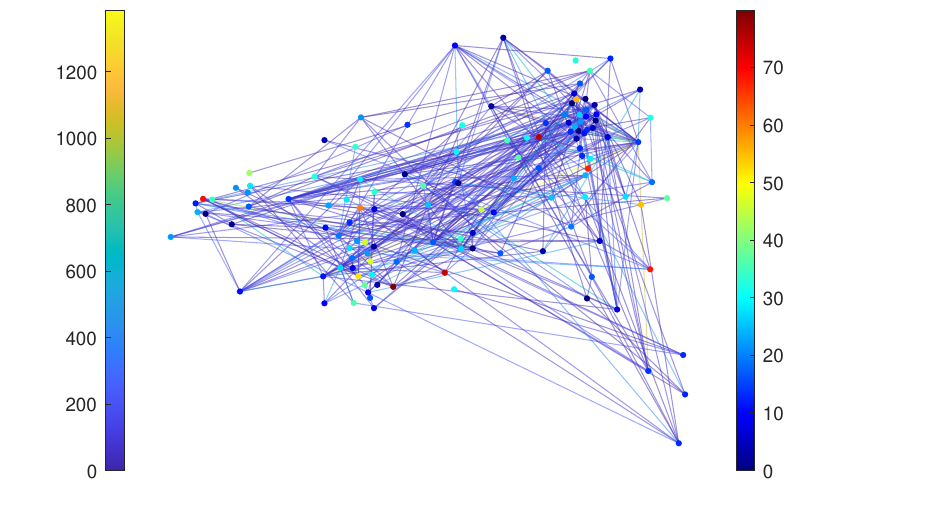}
        \caption{t=31}
        \label{fig:subfig1} 
    \end{subfigure}
    \hfill 
    \begin{subfigure}{0.3\textwidth}
        \includegraphics[trim={100 300 50 300},clip,width=\linewidth]{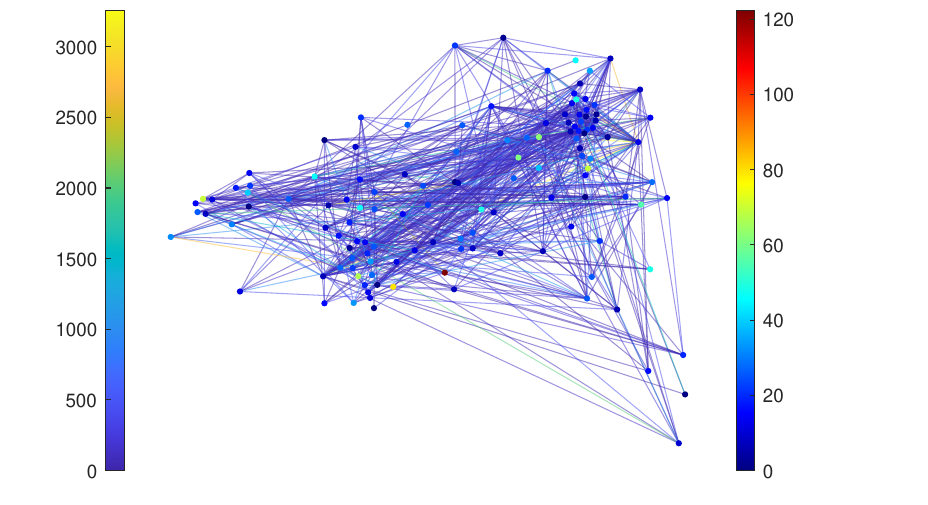}
        \caption{t=41}
        \label{fig:subfig2}
    \end{subfigure}
    \hfill 
    \begin{subfigure}{0.3\textwidth}
        \includegraphics[trim={100 300 50 300},clip,width=\linewidth]{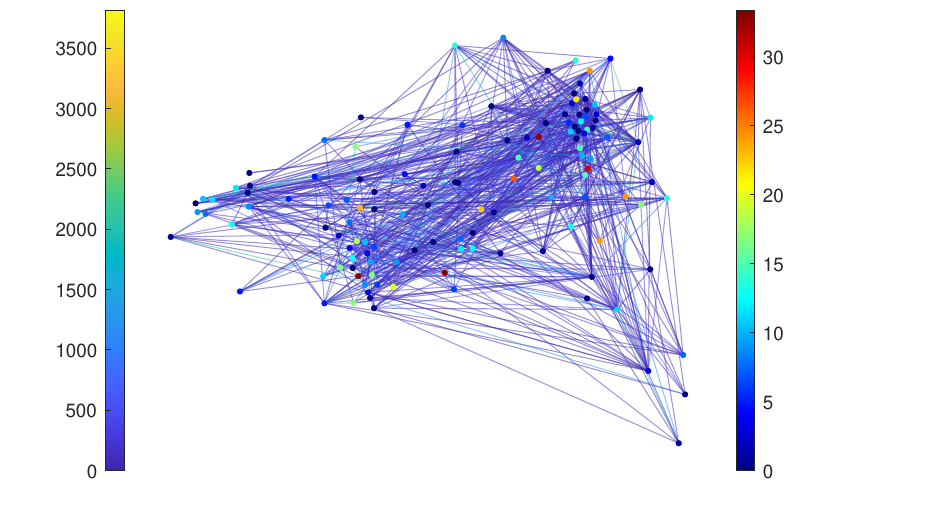}
        \caption{t=61}
        \label{fig:subfig3}
    \end{subfigure}

    \caption{A series of DTDG containing signals on both the nodes (0-simplex) and edges (1-simplex, omitting self-loops), recoding COVID-19 cases on nodes and population mobility on edges. Left color bar: edges. Right color bar: nodes.} %
    \label{fig:pandemic}
\end{figure*}

In the field of DTDG, the following datasets have become widely used, allowing researchers to analyze the evolution of relationships, dependencies, and interactions over discrete time intervals.
\textbf{UN Vote} \cite{bailey2017estimating} captures voting records of countries in the United Nations General Assembly from 1946 onwards, with a node representing a country and an edge between nodes reflecting similar voting behavior in a given year.
\textbf{Elliptic} \cite{weber2019anti} contains Bitcoin transaction data from the Elliptic platform, with each transaction labeled as either "licit"(\eg exchanges, wallet providers, miners) or "illicit"(\eg scams, malware, ransomware). Nodes represent Bitcoin addresses, and edges capture payment flow between these addresses in discrete time steps. 
\textbf{Twitter-Tennis} \cite{beres2018temporal} contains interactions on Twitter related to tennis events, where nodes represent Twitter users and edges represent mentions, replies, or retweets within specific time intervals.
\textbf{ABIDE}~\cite{di2014autism, zhao2024sequential} (Autism Brain Imaging Data Exchange dataset) consists of functional MRI (fMRI) data from individuals diagnosed with autism spectrum disorder and neurotypical controls, which is often used to create brain connectivity networks over time, where nodes represent brain regions and edges represent functional connectivity. As shown in Fig.~\ref{fig:pandemic}, the illustration presents COVID-19 cases at the nodes and population mobility along the edges at various timestamps  \cite{yan2025adaptive}.

\begin{table*}[t]
\centering
\caption{Commonly used datasets in dynamic graphs. * represents the specific number that is not given.}
\resizebox{0.8\textwidth}{!}{
\begin{tabular}{l|llllll}
\toprule
\textbf{Type}          & \textbf{Dataset}     & \textbf{Domain} & \textbf{Category} & \textbf{\#Nodes} & \textbf{\#Edges} & \textbf{\#Timestamps} \\
\midrule
\multirow{12}{*}{STG}  
                       & METR-LA~\cite{li2017diffusion}              & Traffic         & general          & 207              & 1,515            & 34,272                \\
                       & PeMSBay~\cite{li2017diffusion}              & Traffic         & general          & 325              & 2,369            & 52,116                \\                       
                       & Montevideo~\cite{rozemberczki2021pytorch}           & Traffic         & general          & 675              & 690              & 740                   \\
                       & PeMSD7~\cite{yu2018_STGCN}               & Traffic         & general          & 228              & 19,118           & 1,989                 \\
                       & SZ-taxi~\cite{zhao2019tgcn}              & Traffic         & general          & 156              & 532              & 2,977                 \\
                       & Los-loop~\cite{zhao2019tgcn}             & Traffic         & general          & 207              & 2,833            & 2,017                 \\
                       & Traffic~\cite{li2019predicting}              & Traffic         & general          & 4,438            & 8,996            & 2,160              \\
                       & Seattle Loop~\cite{cui2018deep, cui2019traffic} & Traffic 
                           & general          & 323        & 762     & 26,304     \\
                       & PedalMe~\cite{rozemberczki2021pytorch}       & Traffic   & general    & 15     & 225     & *   \\
                       & US Weather~\cite{usweather}    & Meteorological    & general
                       & *  & * & *  \\
                        & Chickenpox~\cite{rozemberczki2021chickenpox}   & Epidemiology   & general  & 20   & 61   & 522       \\
                       & HCP~\cite{van2013wu}   & Neuroimaging       &general    &360    &*  &*  
                       \\
\midrule
\multirow{15}{*}{CTDG} & Social Evolution~\cite{madan2011sensing}     & Proximity       & general          & 74               & 2,099,519        & 565,932               \\
                       & Contact~\cite{sapiezynski2019interaction}              & Proximity       & general          & 692              & 2,426,279        & 8,065                 \\
                       & Bitcoin-Alpha~\cite{kumar2016edge, kumar2018rev2}       & Finance         & general          & 3,783            & 24,186           & 24,186                \\
                       & Bitcoin-OTC~\cite{kumar2016edge, kumar2018rev2}          & Finance         & general          & 5,881            & 35,592           & 35,592                \\
                       & LastFM~\cite{kumar2019predicting}               & Interaction     & bipartite         & 1,980            & 1,293,103        & 1,283,614             \\
                       & MOOC~\cite{kumar2019predicting}                 & Interaction     & bipartite         & 7,144            & 411,749          & 345,600               \\
                       & Enron~\cite{shetty2004enron}                & Social          & general          & 184              & 125,235          & 22,632     \\
                       & UCI~\cite{panzarasa2009patterns}                  & Social          & general          & 1,899            & 59,835           & 58,911             \\
                       & DGraphFin~\cite{huang2022dgraph}            & Social          & general          & 3,700,550        & 4,300,999        & *                     \\
                       & Wikipedia~\cite{kumar2019predicting}            & Social          & bipartite         & 9,227            & 157,474          & 152,757               \\
                       & Reddit~\cite{kumar2019predicting}               & Social          & bipartite         & 10,984           & 672,447          & 669,065               \\
                       & Taobao~\cite{jin2022neural, zhu2018learning}               & E-commerce      & bipartite         & 82,566           & 77,436           & *                     \\
                       & Taobao-Large~\cite{jin2022neural, zhu2018learning}         & E-commerce      & bipartite         & 1,630,453        & 5,008,745        & *                     \\
                       & eBay-Small~\cite{huang2024benchtemp}           & E-commerce      & bipartite         & 38,427           & 384,677          & *                     \\
                       & eBay-Large~\cite{huang2024benchtemp}           & E-commerce      & bipartite         & 1,333,594        & 1,119,454        & *                     \\
    \midrule
\multirow{11}{*}{DTDG} & UN Vote~\cite{bailey2017estimating}              & Politics        & general          & 201              & 1,035,742        & 72                    \\
                       & US Legislative~\cite{huang2020laplacian, fowler2006legislative}       & Politics        & general          & 225              & 60,396           & 12                    \\
                       & Canadian Parliament~\cite{huang2020laplacian}  & Politics        & general          & 734              & 74,478           & 14                    \\
                       & UN Trade~\cite{macdonald2015rethinking}             & Finance         & general          & 255              & 507,497          & 32                    \\
                       & Elliptic~\cite{weber2019anti}             & Finance         & general          & 203,769          & 234,355          & 49                    \\
                       & HEP-TH~\cite{leskovec2005graphs, gehrke2003overview}               & Citation        & general          & 27,770           & 352,807          & 3,487                 \\
                       & MAG~\cite{hu2021ogb, zhou2022tgl}                  & Citation        & general          & 121,751,665      & 1,297,748,926    & 120                   \\
                       & Twitter-Tennis~\cite{beres2018temporal}       & Social          & general          & 1,000            & 40,839           & 120                   \\
                       & Autonomous systems~\cite{leskovec2005graphs}   & Communication   & general          & 7,716            & 13,895           & 733                   \\
                       & Flights~\cite{schafer2014bringing}              & Transport       & general          & 13,169           & 1,927,145        & 122                   \\
                       & ABIDE~\cite{di2014autism, zhao2024sequential}    & Neuroimaging   & general   & 200-360   & *    & *
\\
\bottomrule
\end{tabular}}
\label{tab:datasets}
\end{table*}

\section{Conclusion and Future Direction}
\label{sec_conclusion}
TVGs provide a powerful framework for modeling systems in which both signals and relational structures evolve over time. 
In this review, we examined the connections between TVGSP and TVGNNs, highlighting how theoretical signal-processing principles and data-driven learning models jointly contribute to the analysis of TVG data. 
By extending classical GSP tools to temporal graph settings and combining them with neural architectures capable of capturing nonlinear dependencies, these two paradigms provide complementary approaches for modeling complex real-world systems.

Despite rapid progress, research on TVGs is still in an early stage, and many theoretical and practical challenges remain. 
Continued integration between signal processing theory, machine learning, and emerging data-driven technologies is likely to open new directions for both methodology and applications. 
Building on the developments discussed in this review, several promising research directions are outlined below:

\begin{itemize}

    \item \textbf{Theoretical Foundations and Scalability}: The unified perspective between TVGSP and TVGNNs highlighted in this review provides a foundation for advancing the theoretical understanding of TVG models. 
    Building on this connection, several fundamental questions remain important for the continued development of the field and pushing the frontiers of TVG research, particularly in terms of scalability, stability, convergence, and robustness of TVG-based algorithms.
    As graph sizes and time series lengths grow, scalable algorithms are needed to efficiently handle large-scale TVGs by reducing computational complexity without compromising performance \cite{Li_scalability_2024}.
    Beyond STGs, TVGSP methods still lack mature theoretical formulations and efficient computational frameworks for DTDGs and especially CTDGs, partly due to the need to repeatedly handle topology evolution, irregular event dynamics, and potentially nonlinear dependencies. 
    Developing frameworks that retain spectral interpretability while supporting dynamic and evolving structures remains a key challenge. 
    Incorporating probabilistic or Bayesian frameworks may further enhance representation abilities and uncertainty quantification \cite{das2025bayesian, Rompelberg_Bayesian_2025, Li_2024_alpha}. 
    In parallel, a deeper understanding of phenomena such as over-smoothing and over-squashing in TVGNNs under dynamic graph conditions remains an important topic for future research.

    \item \textbf{Higher-Order Graph Signals}: Beyond node-based representations, many real-world problems require modeling higher-order signals defined not only on nodes, but also on higher-order structures such as edges, faces, and multiway interactions, such as traffic flow on road segments \cite{Jia_2019}, gene interaction dynamics \cite{Kuruoglu_2016_gene}, and multi-agent systems \cite{han_2019_computing}. Higher-order structures, including hypergraphs, simplicial complexes, cell complexes, and multigraphs, provide a natural extension for such representations \cite{Bick_2023_what_are_higher_order}. This has motivated the development of Topological Signal Processing, which generalizes GSP through the Hodge Laplacian to encode signals on simplicial or cell complexes beyond individual nodes \cite{Barbarossa_2020, Roddenberry_2022_SP_cell_complex}. Using these operators, researchers have demonstrated edge-signal smoothing and denoising, embedding techniques for simplicial signals, semi-supervised edge learning, Hodge-wavelet analogues, localized topological Slepians, and FIR/convolutional filtering frameworks \cite{Barbarossa_2020_SPM, martino2019hyper, Schaub_2020_Random, Roddenberry_2022, Battiloro_2022_Topological, Yang_2021_FIR, Sardellitti_2022, Yang_2022_Simplicial, Kadambari_2022_Distributed}. However, these techniques are still rarely explored in the context of STGs and remain largely unexplored for DTDGs and CTDGs. Collectively, these developments indicate that higher-order representations provide a principled pathway for modeling complex relational data, opening opportunities to extend current GSP and GNN approaches toward richer temporal, structural, and multi-entity signal domains \cite{isufi2025topological}.
        
    \item \textbf{Advancing Machine Learning Paradigms for TVGs}: Learning from TVGs themselves remains a relatively new frontier in machine learning. 
    Developing hybrid models that combine the interpretability of TVGSP with broader machine-learning paradigms can unlock new capabilities for learning from TVGs. 
    While TVGNNs provide powerful neural architectures for dynamic graph learning, future research may also explore alternative learning frameworks such as probabilistic models, causal inference methods, and foundation-model–driven TVG analytics. 
    Promising directions include multi-scale and multimodal learning strategies \cite{hu2025adaptive, wang2025chattime}, federated and privacy-preserving learning on TVGs, causal and counterfactual inference on dynamic networks, and machine unlearning for evolving graph systems \cite{Guo_Pan_Zhang_Wang_2025, guo2025spatio}. 
    Beyond deterministic architectures, probabilistic and Bayesian learning frameworks offer promising directions for modeling uncertainty and capturing stochastic dynamics in TVGs. 
    These developments may lead to more robust, generalizable, and explainable learning frameworks for dynamic graph analysis. 

    \item \textbf{LLMs, AIGC, and Dynamic Graph Modeling}: Recent advances in large language models (LLMs) and AI-generated content (AIGC) are beginning to intersect with TVG models, opening new opportunities for both TVGSP and TVGNN research \cite{Qin_Yan_Kuruoglu_2025}. On one hand, LLMs equipped with graph-augmented memory and reasoning modules provide a promising direction for learning high-level temporal and relational structure from dynamic graphs, potentially complementing classical spectral methods with semantic priors and interpretable temporal reasoning. On the other hand, the rapid rise of AIGC introduces new TVG application domains—such as evolving knowledge graphs, interaction networks in multi-agent generative systems, and continually updated recommendation environments—where graph topology and signals evolve at high frequency. These developments motivate hybrid TVG frameworks that integrate principled spectral modeling with data-driven temporal representation learning, and highlight emerging challenges related to scalability, continual adaptation, robustness, and reliability of TVGSP/TVGNN systems in LLM-driven environments.

    \item \textbf{Applications and Real-Time Adaptation}: Exploring novel applications of TVGs in diverse and interdisciplinary domains can significantly propel the field forward. Beyond well-established areas, promising opportunities exist in neuroscience, epidemiology, smart cities, financial systems, cybersecurity, social dynamics, and bioinformatics. These applications benefit from the unique ability of over TVGs to represent and analyze highly dynamic, interconnected systems over time. Additionally, although current TVGSP has shown promising performance for real-time analysis on STGs, few TVGSP and TVGNN approaches extend this to the DTDGs and CTDGs. Future research should develop models capable of quickly adapting to evolving graph structures and signals in real time.  
\end{itemize}

Overall, the integration of TVGSP principles with modern TVGNN frameworks highlights a promising paradigm for modeling and understanding complex time-varying systems. 
As theoretical foundations continue to mature and learning-based methods become more capable of capturing evolving relational structures, the synergy between signal-processing theory and data-driven machine learning is expected to play an increasingly important role in advancing both the methodology and practical applications of TVG analysis.

 




\section*{Acknowledgment}
This work is supported by Shenzhen Science and Technology Innovation Commission Grant JCYJ20220530143002005, Shenzhen Ubiquitous Data Enabling Key Lab Grant ZDSYS20220527171406015, and Tsinghua Shenzhen International Graduate School Start-up Fund QD2022024C.

\bibliography{ref}

\clearpage

\appendix
\section{Graph Acquisition and Representation Techniques}
\label{sec_graph_structure}

The formulation of an appropriate graph topology when TVG signals are present is crucial for the subsequent processing of both TVGSP and TVGNNs. Thus, it is necessary to examine key techniques for representing multivariate signals on topological structures.

The simplest and most common way of acquiring the graph topology $\mathcal{G}$ is when the signals inherently exhibit or form a graph-like structure. 
In such cases, real-world data associated with the nodes can be constructed based on meaningful and realistic relationships.
For example, naturally occurring graph structures appear in traffic forecasting tasks, where sensors placed along highways, recording data like traffic speed or flow, are modeled as nodes, and edges represent their spatial relationships \cite{yu2018_STGCN}.
In the case of meteorological data collection, variables such as longitude and latitude from several weather stations are aggregated and analyzed using a Gaussian Kernel to construct a k-nearest-neighbor graph \cite{Spelta_2020_NLMS}. 
A graph depicting the strength of 5G signal reception is created as detailed in \cite{Lorenzo_2016_GLMS} by documenting signal strength across various sensors. 
A slightly more abstract example of a real-world graph is a social network, where users are represented as nodes and their interactions form edges between them \cite{hamilton2017inductive}.
Another example is a citation network in which each node is a scientific paper and edges represent citations from one paper to another \cite{sen2008collective}. 
While many real-world datasets inherently exhibit graph structures, in other cases, the graph topology may not appear as straightforward.
Specialized methods are employed to construct graphs that capture underlying relationships in the data.

Under the scenario where direct inference of the graph topology is not straightforward, a graph can still be formed.
In statistical graphical models, statistical measures between two data elements are used to define the pairwise relationship. 
For example, the covariance matrix and the precision matrix can be used to form a graph because covariance is a statistical measure of the deviations between two variables, meaning that measuring and properly thresholding this pairwise relationship measure gives us a sparse graph \cite{lauritzen1996graphical}. 
The Gaussian Graphical Model is a specific example of a Statistical Graphical Model that represents a conditional dependency structure in a graph: the nodes of the graph are the multivariate Gaussian distributed variables, and edges between these variables are the conditional dependencies or non-zero partial correlation coefficients \cite{yuan2007_GMM_model, Altenbuchinger_2020_GMM}. 
GSP operations such as graph convolution, transforms, sampling, prediction, and regularization can be realized from a probabilistic and statistical aspect using the Gaussian Graphical Model \cite{zhang2015graph}. 
The graph Laplacian can be learned using a merging of the Gaussian prior and a smoothness assumption on the multivariate signal, which is equivalent to learning the underlying topological structure \cite{dong_2016_learning_laplacian}. 
A multivariate signal can be assumed to be a sparse linear combination of atoms from a dictionary based on a polynomial of graph Laplacians that represents localized patterns from a sparse prior \cite{maretic_2017_graph_sparse_prior}. 
Another sparse prior assumption topology learning approach can be found in \cite{belilovsky_2017_learning_sparse_prior}: the Gaussian Graphical Model is used in a way that data is trained in a neural network to learn a function calculating the empirical covariance matrices to form graph structures. 

Graph embedding techniques leverage adjacency and topological properties to project signals and features onto an embedding space to facilitate the analysis, visualization, and application in downstream GSP tasks. 
In the spectral domain, graph embeddings can be achieved by leveraging the eigenvectors of graph representations such as the adjacency matrix $\mathbf{A}$ \cite{luo_2003_spectral_embedding}. 
In the spatial domain, early approaches to generating the graph embedding are to use the random walk to propagate through the graph and then embed the data on the walk trajectories as words using approaches such as word2vec \cite{mikolov_2013_word_2_vec}.
The DeepWalk generates embeddings by performing random walks on the graph, treating each node sequence like a sentence in natural language processing \cite{perozzi2014deepwalk}. 
The node2vec extended DeepWalk further by introducing a flexible search mechanism that balances breadth-first and depth-first search strategies \cite{grover2016node2vec}. 
Recently, a more common method to generate graph embeddings is through GNNs.
The spectral GCN is achieved by feeding the spectral graph convolution in \eqref{eq_spectral_graph_convolution} into a nonlinear activation function and replacing the filter $\mathbf{H}(\boldsymbol{\Lambda})$ with learnable filters as if the filters are parameters learned from backpropagation \cite{bruna2013_spectral_GCN}. 
The spatial GCN and the ChebNet are the spatial domain versions of GNNs with the spectral convolutions approximated by the Chebyshev polynomials, as shown in \eqref{eq_cheb_H_1}, and let the neural network learn the polynomial coefficient instead of the spectral filter \cite{kipf2017semi_supervised, defferrard2016_cheb}.
GraphSAGE exploits graph aggregations that sample and combine features from node neighborhoods, enabling an inductive framework where the model generalizes better to unseen nodes during inference \cite{hamilton2017inductive}.
The message-passing neural networks (MPNNs) further generalize the previous GNNs by introducing a flexible and localized method in which computations are defined at the node level based purely on the node local adjacencies instead of defining all computations at once \cite{gilmer_2017_neural}. 
These graph embedding techniques establish a foundation for integrating GSP with machine learning tasks, including classification and clustering, by transforming graph signals along with the information in the graph topology into representations suitable for non-graph machine learning models.

\begin{table*}[h]
    \centering
    \caption{Performance comparison of TVGSP and TVGNN models for Wind Speed and Temperature prediction.}
    \label{tab:models}
    \begin{tabular}{c|cc|cc}
        \toprule
        \multirow{2}{*}{Model} & \multicolumn{2}{c|}{Wind Speed} & \multicolumn{2}{c}{Temperature} \\
        \cmidrule{2-3} \cmidrule{4-5}
         & RMSE & MAE & RMSE & MAE \\
        \midrule
        GLMS   & 2.43 ± 0.00 & 1.51 ± 0.00 & 1.30 ± 0.00 & 0.85 ± 0.00 \\
        GNLMS  & 2.97 ± 0.00 & 1.68 ± 0.00 & 1.21 ± 0.00 & 0.77 ± 0.00 \\
        GNS    & 3.03 ± 0.00 & 1.76 ± 0.00 & 1.69 ± 0.00 & 1.13 ± 0.00 \\
        \midrule
        GCN    & 2.52 ± 0.02 & 1.95 ± 0.01 & 2.76 ± 0.05 & 2.02 ± 0.01 \\
        STGCN  & 1.69 ± 0.41 & 1.27 ± 0.45 & 2.55 ± 0.12 & 1.92 ± 0.07 \\
        RGDAN  & 0.77 ± 0.03 & 0.50 ± 0.01 & 1.15 ± 0.10 & 0.79 ± 0.08 \\
    \bottomrule
    \end{tabular}
\end{table*}
The pairwise relationships in graphs can be dynamic.
Time-evolving processes over graphs can be captured by kernels formed by a preselected dictionary that fits the signal of interest, where partial correlation, partial Granger causality, non-linear structural equations, and vector autoregressions techniques are used along with assumptions of the graph signals, such as low-rank, sparsity, and smoothness \cite{giannakis_2018_topology, saboksayr_2021_online_discriminative_graph_learning, Money_2023_Sparse}. 
A dynamic graph can be formed from the correlated properties, which are previously seen in static Statistical Graphical Models, in TVG signals by considering the space-time interactions in the underlying signal \cite{liu2019graph, liu2019spatiotemporal, yamada2019time}. The topology inference on dynamic signals can be approached using a state-space formulation of a network process \cite{coutino_2020_state_space}. 
The Graphical LASSO is another instance of the Statistical Graphical Model method that estimates the inverse covariance matrix (precision matrix) with sparsity constraints enforced through an $l_1$ penalty. 
By combining Graphical LASSO with an ADMM-based algorithm, one can incorporate signals from neighboring time points and pairwise connections to learn the smoothly changing graphs \cite{Yang_2020_estimating_stock_lasso_tv}. 
Another smoothness-based graph topology learning task is to use the Sobolev smoothness as a measure to define a new smoothness function for TVG signals \cite{giraldo_2022_reconstruction_Sobolev_smoothness}. 
The work of Natali et al. proposes to use the structural equation model to learn graph topologies in an online fashion and further proposes a combination of the Gaussian graphical model, the structural equation model, and the smoothness-based model into an all-in-one online topology learning framework \cite{natali2022learning, natali2021_structural_equation_models}. 
The causal relationships between data can be modeled using a Vector Autoregressive model, making it possible to infer a graph topology from the underlying time-series vector \cite{zaman2020online}. 
A specific example of this approach can be seen in a time-varying gene activation graph formed by thresholding a regression matrix of a time-varying gene regression that reflects gene activation patterns over time, leading to a parametric representation of the time-evolution and influence between pairs of data \cite{Kuruoglu_2016_gene}. 
Given an underlying graph structure at a one-time instance, assuming that the dynamics of a TVG only have some minor changes between time instances, a small perturbation analysis of the Laplacian matrix and its GFT can be done to model the time-evolution of the dynamic graph \cite{sardellitti_2021_online_small_pertubation}. 
Topology inference of a dynamic topology from time-varying data can be tackled from another angle, using an iterative prediction-correction approach on a Gaussian Graphical Model \cite{natali2021online}.

\section{Evaluation Metrics}

Different from static GSP, in TVGs, the most popular task is node-level regression rather than classification, as the core in TVGSP is the modeling of graph temporal association. Thus, learning from the past to make accurate predictions of the future is the property that most TVGSP models desire. The metrics for regression are usually drawn from multivariate time series analysis, which mainly include Mean-Square Error (MSE), Mean Absolute Error (MAE)~\cite{trivedi2019dyrep}, root Normalized Mean-Square Error(rNMSE)~\cite{sabbaqi2024inferring}, Root Mean Squared Error (RMSE)~\cite{bagheri2024joint}, and Mean Absolute Percentage Error (MAPE)~\cite{sabbaqi2023graph}. Taking $k^\text{th}$ step ahead prediction as an example, these metrics are defined as functions of node signal $\boldsymbol{x}$ as:
\begin{equation}
    \begin{split}
        \text{MSE}(\boldsymbol{x}) &= \sum_{i=1}^{k}\frac{1}{Nk}\|\hat{\boldsymbol{x}}[t+i] - \boldsymbol{x}[t+i]\|_2^2,\\
        \text{MAE}(\boldsymbol{x}) &= \sum_{i=1}^{k}\frac{1}{Nk}\|\hat{\boldsymbol{x}}[t+i]- \boldsymbol{x}[t+i]\|_1,\\
        \text{RMSE}(\boldsymbol{x}) &= \sqrt{\sum_{i=1}^{k}\frac{1}{Nk} \|\hat{\boldsymbol{x}}[t+i] - \boldsymbol{x}[t+i]\|_2^2},\\
        \text{rNMSE}(\boldsymbol{x}) &= \sum_{i=1}^{k}\frac{\|\boldsymbol{\hat{x}}[t+i] - \boldsymbol{x}[t+i]\|_2}{k\times\|\boldsymbol{x}[t+i]\|_2},\\
        \text{MAPE}(\boldsymbol{x}) &= \sum_{i=1}^{k}\frac{\|\boldsymbol{\hat{x}}[t+i] - \boldsymbol{x}[t+i]\|_1}{{Nk}\times\|\boldsymbol{x}[t+i]\|_1}.
    \end{split}
    \label{eq:metric}
\end{equation}
In GSP and graph machine learning algorithms, the Frobenius norm serves as an alternative to the $l_2$ norm, where functions are constructed between the node signal matrix $\mathbf{X}$ and the estimated node signal matrix $\hat{\mathbf{X}}$. In addition to regression, there are also classification tasks on TVGs that usually use Accuracy, Precision, Recall, Area Under Curve for binary classification, and micro-F1 and macro-F1 for multi-class classification~\cite{kumar2019JODIE, trivedi2019dyrep, rossi2020temporal, xu2020inductive}. The formulation of these metrics is the same as in static graph signal classification and conventional machine learning tasks. Table~\ref{tab:models} illustrates the performance of various TVGSP models on a 1-step ahead spatio-temporal prediction task.
The methods being compared are from 2 categories: online GSP methods GLMS~\cite{Lorenzo_2016_GLMS}, GNLMS~\cite{Spelta_2020_NLMS}, and GNS~\cite{peng2023adaptive}, as well as offline TVGNN models such as STGCN~\cite{yu2018_STGCN}, and RGDAN~\cite{fan2024rgdan}. 
We included the GCN~\cite{kipf2017semi_supervised} as the non-temporal GNN baseline.
These models are evaluated for wind speed and temperature prediction using the RMSE and MAE.

Besides node-level tasks, there are also some edge-level and graph-level tasks on TVGs (\eg edge prediction and graph classification). Note that, in an unweighted graph, the edge prediction can be viewed as edge classification (\ie to judge whether the edge exists in the future or not). Thus, the metrics that are popular in classification can also be utilized for them~\cite{zhang2022dynamic, zhao2024sequential, fu2021sdg}. Besides, for graph topology regression, the regression error of the estimated graph topology can be computed by comparing the graph adjacency matrix. In this situation, the input to the regression metrics functions defined in (\ref{eq:metric}) becomes the adjacency matrix, and the L2 norm becomes the Frobenius norm~\cite{peng2021dynamic}. 

In graph lifelong learning, the evaluation metrics change slightly, as its goal is to maintain the ability to make predictions on previous nodes/classes. In this way, the metrics that are usually used include "AP" and "AF," which represent average performance (\eg the average classification accuracy or regression error) and average forgetting rate. They can be computed by:
 \begin{equation}
     \begin{split}
         \text{AP} &= \frac{1}{K}\sum_i^K a_i, \\
         \text{AF} &= \frac{1}{K-1} \sum_i^K f_i^K,\\
         f_i^K &= \max_{k\in \{i,\cdots, K-1\}} a_{k,i} - a_{K,i},
     \end{split}
 \end{equation}
where $a_i$ denotes the accuracy of task $i$; $a_{j,i}$ represents the accuracy of task (or classes) $i$ after the model being trained on task $j$; $K$ is the total number of tasks; $f$ represents the forgetting rate.

\begin{table}[h]
\centering
\caption{Summary of notations used in this paper. $[t]$ denotes the time index.}
\begin{tabular}{ll}
\toprule
    Notation & Meaning \\
    \midrule
    $\mathcal{G}$ & Graph \\
    $\mathcal{V}$ & Node set \\
    $\mathcal{E}$ & Edge set \\
    $N$ & Number of nodes \\
    $\boldsymbol{x}[t]$ & Graph signal at time $t$ \\
    $\mathbf{A}$ & Adjacency matrix \\
    $\mathbf{L}$ & Laplacian matrix  \\
    $\mathbf{U}$ & Eigenvector matrix of $\mathbf{L}$\\
    $\Lambda$ & Eigenvalue matrix of $\mathbf{L}$\\
    $\mathbf{H}(\Lambda)$ & Spectral filter \\
    $\mathbf{W} $ & Weights \\
    $\mathbf{D}_S$ & Sampling / masking matrix \\
    $\mathcal{L}$ & Loss function \\
    $\Delta$ & Temporal Change \\
    TVG & Time-Varying Graph \\
    STG & Spatiotemporal Graph \\
    DTDG & Discrete-Time Dynamic Graph \\
    CTDG & Continuous-Time Dynamic Graph \\
    STGNN & Spatiotemporal Graph Neural Network \\
    DGNN & Dynamic Graph Neural Network \\
    GFT & Graph Fourier Transform \\
    JFT & Joint Time–Graph Fourier Transform \\
\bottomrule
\end{tabular}
\label{tab:notation_final}
\end{table}



\end{document}